\documentclass[preprintnumbers,floatfix,letterpaper,aps,prd,nofootinbib,
twocolumn
]{revtex4-1}
\usepackage{amsmath,amsfonts,latexsym,amssymb,graphicx,graphics,epsfig,subfigure,color,makeidx}
\usepackage{xcolor,diagbox}
\usepackage{multirow}
\usepackage[colorlinks,linkcolor=blue,anchorcolor=blue,citecolor=blue,urlcolor=blue]{hyperref}
\usepackage{mathrsfs}
\usepackage{amsmath}
\usepackage{bm,graphicx,dcolumn,epstopdf,epsf,latexsym,mathbbol, amssymb,amsmath,color,slashed, mathrsfs,mathcomp,simplewick}
\usepackage{makecell}

\newcommand{\bq}{\begin{equation}}
\newcommand{\eq}{\end{equation}}
\newcommand{\bqn}{\begin{eqnarray}}
\newcommand{\eqn}{\end{eqnarray}}
\newcommand{\nb}{\nonumber}
\newcommand{\lb}{\label}
\newcommand{\f}{\frac} 
\newcommand{\p}{\partial}
\newcommand{\tx}{\text}

\newcommand{\lf}{\left}
\newcommand{\rt}{\right}
\begin{document}

\title{Parameterized quasinormal frequencies and Hawking radiation for axial gravitational perturbations of a holonomy-corrected black hole}

\author{Sen Yang${}^{a, b}$}
\email{120220908881@lzu.edu.cn}

\author{Wen-Di Guo${}^{a, b}$}
\email{guowd@lzu.edu.cn}

\author{Qin Tan${}^{c}$}
\email{tanqin@hunnu.edu.cn}

\author{Li Zhao${}^{a, b}$}
\email{lizhao@lzu.edu.cn}

\author{Yu-Xiao Liu${}^{a, b}$}
\email{liuyx@lzu.edu.cn}

\affiliation{
	${}^{a}$ Lanzhou Center for Theoretical Physics,\\ Key Laboratory for Quantum Theory and Applications of the Ministry of Education,\\
    Key Laboratory of Theoretical Physics of Gansu Province,\\
    School of Physical Science and Technology,\\
    Lanzhou University, Lanzhou 730000, China\\
	${}^{b}$ Institute of Theoretical Physics $\&$ Research
	Center of Gravitation, Lanzhou University, Lanzhou 730000, China\\
	${}^{c}$ Department of Physics, Key Laboratory of Low Dimensional Quantum Structures and Quantum Control of Ministry of Education, Synergetic Innovation Center for Quantum Effects and Applications, Hunan Normal University, Changsha, 410081, Hunan, China\\}
\date{\today}

\begin{abstract}
	
As the fingerprints of black holes, quasinormal modes are closely associated with many properties of black holes. Especially, the ringdown phase of gravitational waveforms from the merger of compact binary components can be described by quasinormal modes. Serving as a model-independent approach, the framework of parameterized quasinormal frequencies offers a universal method for investigating quasinormal modes of diverse black holes. In this work, we first obtain the Schr\"{o}dinger-like master equation of the axial gravitational perturbation of a holonomy-corrected black hole. We calculate the corresponding quasinormal frequencies using the Wentzel-Kramers-Brillouin approximation and asymptotic iteration methods. We investigate the numerical evolution of an initial wave packet on the background spacetime. Then, we deduce the parameterized expression of the quasinormal frequencies and find that $r_0 \leq 10^{-2}$ is a necessary condition for the parameterized approximation to be valid, where $r_0$ is the 
quantum parameter. We also study the impact of the parameter $r_0$ on the greybody factor and Hawking radiation. With more ringdown signals of gravitational waves detected in the future, our research will contribute to the study of the quantum properties of black holes.
\end{abstract}

%\pacs{ 04.50.-h, 11.27.+d}

\maketitle

\section{Introduction}
\label{Introduction}
\renewcommand{\theequation}{1.\arabic{equation}} 
\setcounter{equation}{0}

Einstein's general relativity predicts the existence of gravitational waves, a phenomenon first directly observed by LIGO/Virgo in 2015 \cite{LIGOScientific:2016aoc}. Subsequently, gravitational waves have emerged as a novel avenue for investigating physics and astronomy \cite{LIGOScientific:2016lio}. The LIGO–Virgo–KAGRA collaboration completed three observing runs, identifying over 90 significant gravitational wave events from the merger of binary compact objects \cite{LIGOScientific:2018mvr, LIGOScientific:2020ibl, LIGOScientific:2021usb, LIGOScientific:2021djp}. Presently, this collaboration has commenced its fourth observing run. These detected gravitational wave signals provide opportunities to test general relativity within strong gravitational fields \cite{LIGOScientific:2019fpa, LIGOScientific:2020tif, LIGOScientific:2021sio}. The gravitational waveform resulting from the merger of binary compact objects encompasses three phases: inspiral, merger, and ringdown. During the ringdown phase, the previously unstable post-merger black hole stabilizes due to gravitational radiation. This stabilization process can be described by the black hole perturbation theory \cite{Chandrasekhar:1985kt, M. Maggiore}.

Quasinormal modes (QNMs) represent the eigenmodes of a perturbed black hole system. As the distinctive signature of a black hole, QNMs possess complex frequencies that are solely determined by the properties of a 
black hole \cite{Kokkotas:1999bd, Nollert:1999ji, Berti:2009kk, Konoplya:2011qq}. The investigation of QNMs within black holes in general relativity was pioneered by Regge \cite{Regge:1957td}, Wheeler, Zerilli \cite{Zerilli:1970se}, Moncrief \cite{Moncrief:1974gw, Moncrief:1974ng}, and Teukolsky \cite{Teukolsky:1972my}. Recently, a gauge-invariant approach for analytically computing metric perturbations in general spherically symmetric spacetimes was developed in Ref.~\cite{Liu:2022csl}. In the context of QNMs of black holes, the primary objective is to compute numerical results for the quasinormal frequencies. Across the evolution of black hole perturbation theory, various numerical techniques have emerged, including the Wentzel-Kramers-Brillouin (WKB) approximations \cite{Schutz:1985km, Iyer:1986np, Konoplya:2003ii, Matyjasek:2017psv, Konoplya:2019hlu}, Leaver's continued fraction method \cite{Leaver:1985ax, Leaver:1990zz}, the asymptotic iteration method \cite{AIM, AIM2, Cho:2009cj, Cho:2011sf}, and so on \cite{Motl:2003cd, Horowitz:1999jd, 
Berti:2009wx, Lin:2016sch, Shen:2022xdp}.

Indeed, for four-dimensional spherically symmetric black holes in various gravitational theories, the procedure for computing quasinormal frequencies is consistent. This process can be divided into two steps: firstly, deriving the master equation governing linear perturbations, and secondly, employing various numerical methods to solve the master equation. To investigate the influence of the environment, a perturbative formula for black hole QNMs was initially proposed in Refs. \cite{Leung:1997was, Leung:1999iq}. The core of the perturbative formula is treating the influence of the environment as a small perturbation to the effective potential in the master equation. Then, the framework of parameterized quasinormal frequencies was developed in Refs. \cite{Cardoso:2019mqo, McManus:2019ulj} by treating the effects of modified gravitational theories as corrections to the metric and the effective potential for perturbations of Schwarzschild black holes in general relativity. As an effective approach for addressing the QNM problem in black holes, the framework of a parameterized quasinormal frequency has now been extensively studied \cite{Kimura:2020mrh, Hatsuda:2020egs, Churilova:2019jqx, Volkel:2022aca, Hatsuda:2023geo,
Franchini:2022axs,Hirano:2024fgp, Bamber:2021knr}. Note that the parametrized QNM framework provides a novel avenue for constraining modified gravitational theories beyond general relativity \cite{Volkel:2022khh}.

Despite successfully passing numerous astrophysical tests, general relativity is widely acknowledged as incomplete, primarily due to spacetime singularities. Loop quantum gravity emerges as a framework endeavoring to reconcile quantum mechanics with gravity \cite{Ashtekar:2004eh}. Within loop quantum gravity, black hole solutions exhibit regularity \cite{Perez:2017cmj, Bojowald:2020dkb, Ashtekar:2023cod, Zhang:2023yps}. Recent research \cite{Papanikolaou:2023crz} delves into the formation of primordial black holes within this framework, revealing a significant abundance of very small mass black holes compared to predictions under general relativity. QNMs of black holes in loop quantum gravity have been extensively investigated \cite{Santos:2015gja, Liu:2020ola, Daghigh:2020mog, Daghigh:2020fmw, Bouhmadi-Lopez:2020oia, Santos:2021wsw, Fu:2023drp, Moreira:2023cxy, Bolokhov:2023bwm, Yang:2023gas}. Notably, a holonomy-corrected black hole in loop quantum gravity was proposed in Refs.~\cite{Alonso-Bardaji:2021yls, Alonso-Bardaji:2022ear}, by integrating anomaly-free holonomy corrections via a canonical transformation of the general relativity Hamiltonian. Some properties of this black hole have already been investigated in some works, such as scalar and electromagnetic field perturbations \cite{Fu:2023drp, Moreira:2023cxy, Bolokhov:2023bwm}, the gravitational lens effect \cite{Junior:2023xgl, Soares:2023uup}, and so on \cite{Chen:2023bao, Gingrich:2024tuf, Balart:2024rts}.

In this work, we focus on the axial gravitational perturbation of the holonomy-corrected black hole. Within the framework of loop quantum gravity, the quantum effect can be represented by an anisotropic perfect fluid \cite{Ashtekar:2023cod}. Assuming a description wherein the holonomy-corrected black hole is governed by Einstein's gravity minimally coupled with an anisotropic perfect fluid, we derive the master equation governing its axial gravitational perturbation. Utilizing the WKB approximation and asymptotic iteration methods, we compute the corresponding quasinormal frequencies and investigate the influence of the quantum parameter on these frequencies. We explore the numerical evolution of gravitational perturbations on the holonomy-corrected black hole spacetime by assuming a Gaussian packet perturbation. Using the framework proposed in Ref. \cite{Cardoso:2019mqo}, we derive the parameterized expression for quasinormal frequencies for the axial gravitational perturbations of the holonomy-corrected black hole. We investigate the applicability conditions of the parameterized quasinormal frequencies method for the axial gravitational perturbations of the holonomy-corrected black hole. Finally, we investigate the influence of the quantum correction on the greybody factor and  Hawking radiation.

This paper is organized as follows. In Sec. \ref{sec2}, we derive the master equation of the axial gravitational perturbation of the holonomy-corrected black hole.  We calculate the corresponding quasinormal frequencies with the WKB approximation and the asymptotic iteration method. We also study the numerical evolution of an initial Gaussian wave packet under the effective potential in the master equation. In Sec.~\ref{sec3}, we derive the parameterized
expression for quasinormal frequencies for the axial gravitational perturbations of the holonomy-corrected black
hole and explore the applicability conditions for the parameterized
expression. Then we analyze the influence of the quantum correction on the greybody factor and Hawking radiation in Sec. \ref{sec4}. Finally, the conclusions and discussions of this work are given in Sec. \ref{sec5}. Throughout the paper, we use the geometrized unit system with $G = c = 1$.

\section{axial Gravitational perturbation of the holonomy-corrected black hole} 
\label{sec2}
\renewcommand{\theequation}{2.\arabic{equation}} 
\setcounter{equation}{0}

The metric of the holonomy-corrected black
hole in loop quantum gravity proposed in Refs. \cite{Alonso-Bardaji:2021yls, Alonso-Bardaji:2022ear} is  
\bq\lb{metric}
ds^2 = - f(r) dt^2 + \frac{dr^2}{g(r) f(r)} + r^2 \left( d \theta^2 + \sin ^2 \theta d \varphi ^2 \right) ,
\eq 
where the functions $f(r)$ and $g(r)$ are
\bqn
f(r) = 1 - \f{2M}{r},~~~~ 
g(r) = 1 - \f{r_0}{r}.
\eqn
The quantum parameter $r_0$ is defined by
\bqn\lb{quantum parameter}
r_0 = 2M \f{{\lambda}^2}{1 + {\lambda}^2},
\eqn 
where $M$ is a constant of motion, and $\lambda$ is a dimensionless constant that related to the fiducial length of the holonomies. It is worthwhile to mention that $r_0$ defines a minimal spacelike hypersurface, whose area is $4 \pi r_0^2$, separating the trapped black hole interior from the anti-trapped white hole region. The Komar, Arnowitt–Deser–Misner, and Misner-Sharp masses of the holonomy-corrected Schwarzschild black
hole are given by \cite{Alonso-Bardaji:2022ear}
\bqn
M_\text{K} &=& M \sqrt{1-\f{r_0}{r}},~~~~M_\text{ADM} = M + \f{r_0}{2}, \nb\\
M_\text{MS} &=& M + \f{r_0}{2}-\f{M r_0}{r}.
\eqn
We have $M>0$ and $0 < r_0 < 2M$ for an astrophysical black hole. The event horizon of the holonomy-corrected Schwarzschild black
hole is $r_\text{H} = 2M$. When $\lambda \rightarrow 0$, $r_0 \rightarrow 0$, and the metric (\ref{metric}) goes back to the case of the Schwarzschild black hole.

\begin{figure}[!t]
	\centering
	\includegraphics[scale =0.28]{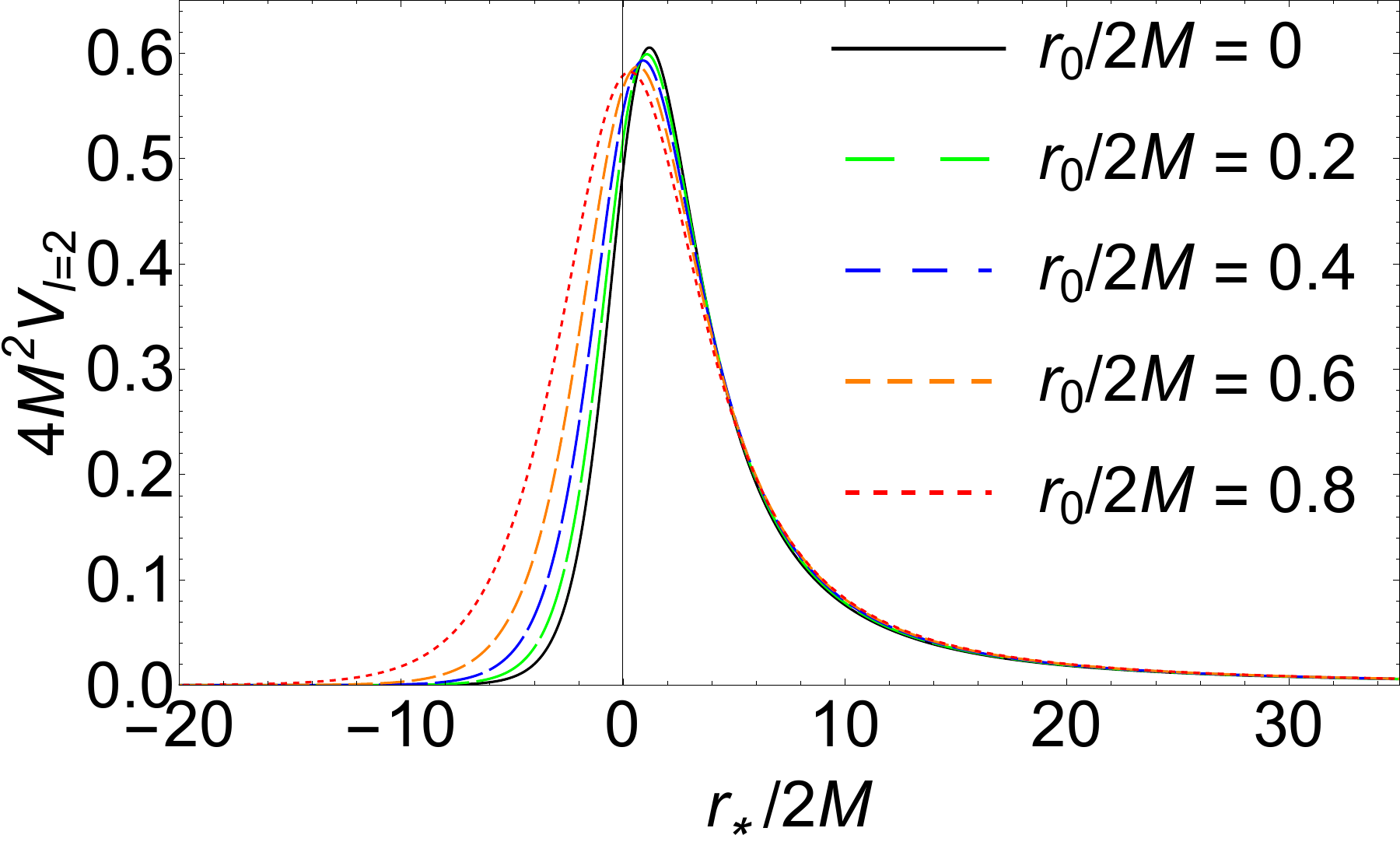}
	\includegraphics[scale =0.28]{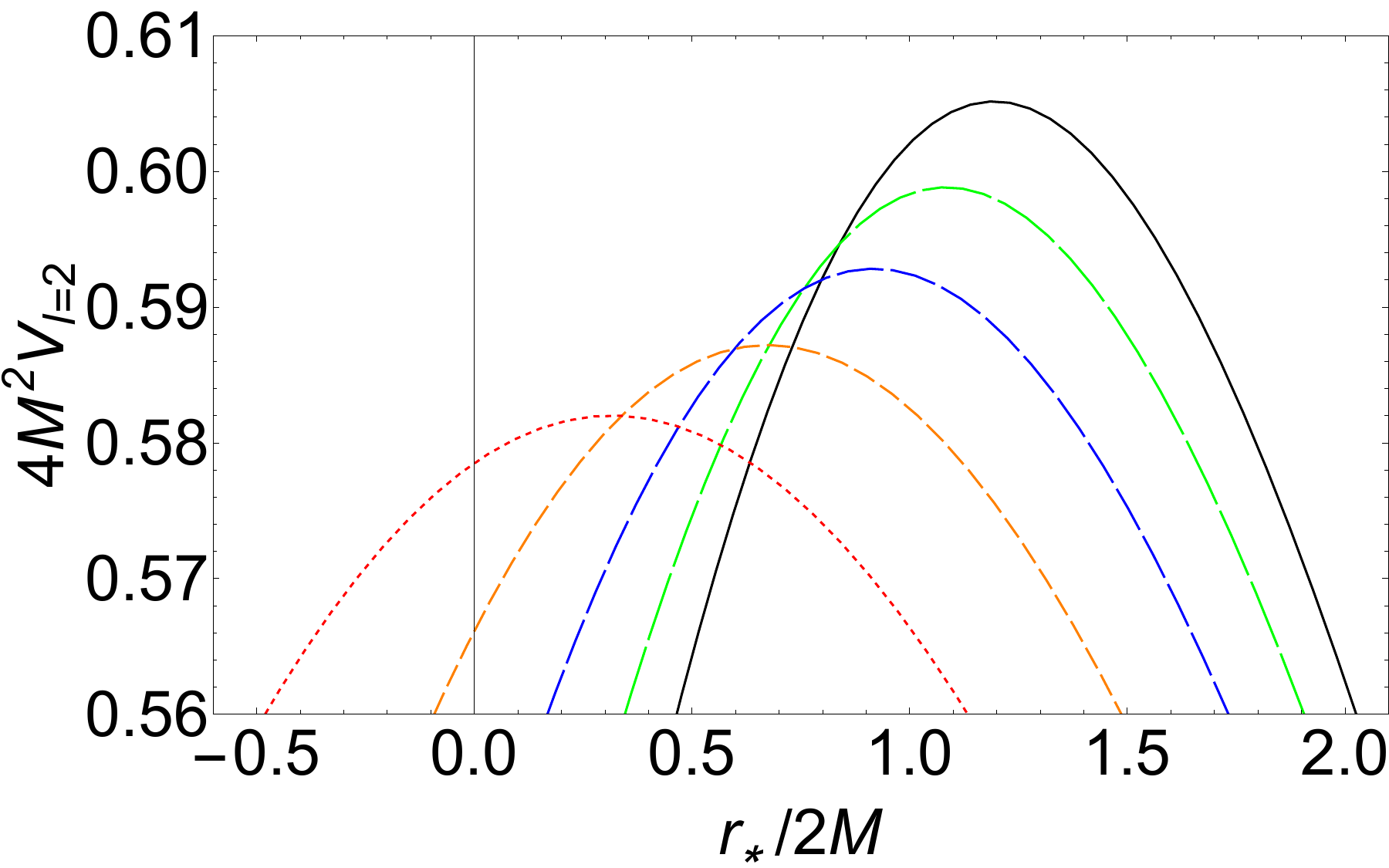}
	\caption{The effective potential (\ref{V}) in the tortoise coordinate (\ref{tortoise}) with $M = 1/2$, $l = 2$, and different value for the parameter $r_0$. The black curve shows the Regge–Wheeler potential of the Schwarzschild spacetime.}
	\label{plot-V}
\end{figure}

Using the method developed in Ref.~\cite{Chen:2019iuo}, we obtain the Schr\"{o}dinger-like master equation of the axial gravitational perturbation of the holonomy-corrected Schwarzschild black hole:
\bqn\lb{master-eq}
\frac{d^2 \Psi}{d r^2_{\ast}} + \lf[ \omega^2 -V(r) \rt] \Psi =0,
\eqn 
where the effective potential is
\bqn\lb{V}
V (r) = &&\frac{f(r)(l-1)(l+2)}{r^2} + \f{2 g(r) f^2(r) }{r^2} \nb\\
&&-  \f{f(r)\sqrt{g(r)}}{r}\frac{d}{dr} \left( f(r) \sqrt{g(r)} \right) ,
\eqn
and $r_\ast$ is the tortoise coordinate defined by
\bqn\lb{tortoise}
r_{\ast} &=& \int \frac{dr}{f(r) \sqrt{ g(r)}} .
\eqn
The effective potential \eqref{V} we derived is consistent with that obtained in Ref.~\cite{Gingrich:2024tuf}. We plot the effective potential \eqref{V} in the tortoise coordinate \eqref{tortoise} with different values of the parameter $r_0$ in Fig.~\ref{plot-V}. From Fig.~\ref{plot-V}, one can see that the effective potential \eqref{V} is a barrier, and goes to zero at both $r_\ast \rightarrow -\infty$, which corresponds $r$ goes to the event horizon, and $r_\ast \rightarrow \infty$, which corresponds $r$ goes to the spatial infinity. And Fig. \ref{plot-V} shows that the height of the effective potential decreases, and the position of the peak of the effective potential moves towards the event horizon with the parameter $r_0$. The influence of the parameter $r_0$ on the height of the effective potential here agrees with the case of the massless scalar field perturbations of the holonomy-corrected Schwarzschild black hole \cite{Fu:2023drp, Moreira:2023cxy}, and differs from the case of the electromagnetic field perturbation \cite{Fu:2023drp}.

For the black hole perturbation problem here, we assume that the two physical boundary conditions are the ingoing wave at the event horizon and the outgoing wave at spatial infinity. Setting $M = 1/2$, $r_0$ from $0$ to $0.9$, and ${l = 2, 3, 4}$, we calculate the quasinormal frequencies of the $n = 0$ modes of the axial gravitational perturbation of the holonomy-corrected Schwarzschild black hole by the 6th-order WKB approximation method \cite{Konoplya:2019hlu} and the asymptotic iteration method with 30th-order expansion \cite{Cho:2009cj, Cho:2011sf}. The details of the WKB approximation and the asymptotic iteration method employed in this work are documented in the appendix \ref{AppA} and appendix \ref{AppB}, respectively. The results of quasinormal frequencies are listed in Tab.~\ref{T-QNFs}. We calculate the numerical errors of the quasinormal frequencies obtained using the 6th-order WKB approximation, as detailed in Eq.~\eqref{WKB-err}, and using the asymptotic iteration method with 30th-order expansion, as outlined in Eq.~\eqref{AIM-err}, respectively.  We also
calculate the relative errors between quasinormal frequencies obtained from the 6th-order WKB approximation and the asymptotic iteration method with 30th-order expansion defined by
\bq
\Delta_\text{WA} = \left| \f{\omega_{\text{WKB}} - \omega_{\text{AIM}}}{\omega_{\text{WKB}}} \right|.
\eq 
The results of the numerical errors are listed in Tab.~\ref{T-errs}. It shows that the numerical errors of the quasinormal frequencies calculated from both the WKB approximation and the asymptotic iteration method are very small, and the quasinormal frequencies calculated from the two methods agree well with each other. And the quasinormal frequencies in Tab.~\ref{T-QNFs} exhibit a high degree of agreement with those obtained in Ref. \cite{Gingrich:2024tuf}. With $r_0$ varying from $0$ to $0.8$, it is shown that the values of the real parts of the quasinormal frequencies increase and the absolute values of the imaginary parts of the quasinormal frequencies decrease. In fact, the constraints from solar system tests on the parameters in Eq. \eqref{quantum parameter} show that $r_0 < 0.8$ \cite{Chen:2023bao}.

\begin{table*}[t]
	\renewcommand\arraystretch{1.5} 
	\centering
	\begin{tabular}{|c|c|c|c|c|c|c|}%{|c{1.0cm}|c{3.0cm}|c{2.0cm}|c{3.0cm}|c{2.0cm}|}
		\hline
		\multicolumn{2}{|c|}{$r_0 /2M$} &$ 0 $ & $ 0.1 $ &$ 0.2 $ & $ 0.3 $ & $ 0.4 $ \\
		\hline
		\multirow{2}*{$ 2 M \omega_{02} $} & WKB & $ 0.747239 - 0.177782 i$ & $ 0.747608 - 0.171702 i$ & $ 0.747856 - 0.165458 i$ & $ 0.748028 - 0.158970 i$ & $ 0.748179 - 0.152158 i$ \\
		~ &AIM & $ 0.747343 - 0.177928 i$ & $ 0.747595 - 0.171800i$ & $ 0.747839 -0.165463i$ & $ 0.748067 -0.158887i$ & $ 0.748271 -0.152036i$ \\
		\hline
        \multirow{2}*{$ 2 M \omega_{03} $} & WKB & $ 1.198890 - 0.185405 i$ & $ 1.199380 - 0.179234 i$ & $ 1.199860 - 0.172822 i$ & $ 1.200340 - 0.166141i$ & $ 1.200790 - 0.159155 i$ \\
		~ & AIM & $ 1.198890 - 0.185406 i$ & $ 1.199380 - 0.179233i$ & $ 1.199860 -0.172820i$ & $ 1.200340 -0.166138i$ & $ 1.200790 -0.159152i$ \\
        \hline
		\multirow{2}*{$ 2 M \omega_{04} $} & WKB & $ 1.618360 - 0.188328 i$ & $ 1.618830 - 0.182039i$ & $ 1.619300 - 0.175507 i$ & $ 1.619750 - 0.168704  i$ & $ 1.620170 - 0.161594 i$ \\
		~ & AIM & $ 1.618360 - 0.188328 i$ & $1.618830 - 0.182038 i$ & $ 1.619300 - 0.175506i$ & $ 1.619750 -0.168703i$ & $ 1.620170 -0.161593i$ \\
		\hline 
		\multicolumn{2}{|c|}{$r_0 /2M$} &$ 0.5 $ & $ 0.6 $ &$ 0.7 $ & $ 0.8 $ & $ 0.9 $ \\
		\hline
		\multirow{2}*{$ 2 M \omega_{02} $} & WKB & $ 0.748530 - 0.144954 i$ & $ 0.748558 - 0.137306 i$ & $ 0.748685 - 0.129177i$ & $ 0.748642 - 0.120539 i$ & $ 0.748214 - 0.111334 i$ \\
		~ &AIM & $ 0.748435 - 0.144864 i$ & $ 0.748536 - 0.137314 i$ & $ 0.748536 - 0.129313 i$ & $ 0.748371 - 0.120767i$ & $ 0.747847 - 0.111449 i$ \\
		\hline
		\multirow{2}*{$ 2 M \omega_{03} $} & WKB & $ 1.201200 - 0.151819 i$ & $ 1.201570 - 0.144079 i$ & $ 1.201850 - 0.135863 i$ & $ 1.202010 - 0.127081  i$ & $ 1.201980 - 0.117612 i$ \\
		~ & AIM & $ 1.201200 - 0.151817 i$ & $ 1.201570 - 0.144077 i$ & $ 1.201850 - 0.135862 i$ & $ 1.202010 - 0.127079i$ & $ 1.201980 - 0.117616 i$\\
		\hline
        \multirow{2}*{$ 2 M \omega_{04} $} & WKB & $ 1.620570 - 0.154134 i$ & $ 1.620920 - 0.146270 i$ & $ 1.621210 - 0.137932 i$ & $ 1.621390 - 0.129030  i$ & $ 1.621420 - 0.119443i$ \\
		~ & AIM &  $ 1.620570 - 0.154133 i$ & $ 1.620920 - 0.146269 i$ & $ 1.621200 - 0.137931 i$ & $ 1.621390 - 0.129029 i$ & $ 1.621420 - 0.119442 i$ \\
		\hline
	\end{tabular}
	\caption{The quasinormal frequencies of the $n=0$ modes for the axial gravitational perturbation of the holonomy-corrected Schwarzschild black hole with different values of $r_0$ and $l = 2,~3,~4$, calculated by the WKB approximation method and the asymptotic iteration method.}
	\label{T-QNFs}
\end{table*}

\begin{table*}[t]
	\renewcommand\arraystretch{1.5} 
	\centering

	\begin{tabular}{|c|c|c|c|c|c|c|}%{|c{1.0cm}|c{3.0cm}|c{2.0cm}|c{3.0cm}|c{2.0cm}|}
		\hline
		\multicolumn{2}{|c|}{$r_0 /2M$} &$ 0 $ & $ 0.1 $ &$ 0.2 $ & $ 0.3 $ & $ 0.4 $ \\
		\hline
		\multirow{2}*{$ 2 M \omega_{02} $} & $\Delta_\text{WKB}$ & $ 1.47\times 10^{-4}  $ & $ 7.99 \times 10^{-5} $ & $ 1.97 \times 10^{-5} $ & $ 9.32 \times 10^{-5} $ & $ 1.73 \times 10^{-4} $ \\
		~ & $\Delta_\text{AIM}$ & $ 4.93 \times 10^{-6}$ & $ 4.96 \times 10^{-6}$ & $ 5.02\times 10^{-6}$ & $ 5.14\times 10^{-6}$ & $ 5.33\times 10^{-6}$ \\
        ~ & $\Delta_\text{WA}$ &  $ 0.023320 \% $ & $ 0.012890 \% $ & $ 0.002309\% $ & $ 0.011970\% $ & $ 0.020010\% $ \\
		\hline
        \multirow{2}*{$ 2 M \omega_{03} $} & $\Delta_\text{WKB}$ & $ 2.15 \times 10^{-6} $ & $ 1.50 \times 10^{-6} $ & $ 3.62 \times 10^{-6}$ & $5.00 \times 10^{-6} $ & $ 5.06 \times 10^{-6} $ \\
		~ & $\Delta_\text{AIM}$ &  $ 7.20\times 10^{-7} $ & $ 7.21\times 10^{-7} $ & $ 7.32\times 10^{-7} $ & $ 7.57\times 10^{-7}$ & $ 8.05\times 10^{-7}$ \\
        ~ & $\Delta_\text{WA}$ &  $ 0.000117\%$ & $ 0.000095\% $ & $ 0.000204\%$ & $ 0.000232\%$ & $  0.000208\%$ \\
        \hline
		\multirow{2}*{$ 2 M \omega_{04} $} & $\Delta_\text{WKB}$ & $ 1.87 \times 10^{-7} $ & $ 6.64\times10^{-7} $ & $ 9.83\times10^{-7} $ & $ 1.16 \times 10^{-6} $ & $ 1.23 \times 10^{-6} $ \\
		~ & $\Delta_\text{AIM}$ &  $ 1.81\times 10^{-7}$ & $ 1.83\times 10^{-7}$ & $1.88\times 10^{-7}$ & $1.99\times 10^{-7}$ & $2.17\times 10^{-7}$ \\
        ~ & $\Delta_\text{WA}$ &  $ 0.000010\%$ & $ 0.000022\% $ & $ 0.000029\%$ & $ 0.000031\%$ & $ 0.000031\% $ \\
		\hline 
		\multicolumn{2}{|c|}{$r_0 /2M$} &$ 0.5 $ & $ 0.6 $ &$ 0.7 $ & $ 0.8 $ & $ 0.9 $ \\
		\hline
		\multirow{2}*{$ 2 M \omega_{02} $} & $\Delta_\text{WKB}$ & $1.55 \times 10^{-4} $ & $ 5.36 \times 10^{-4} $ & $ 3.71 \times 10^{-4} $ & $ 1.69 \times 10^{-3}$ & $ 6.80 \times 10^{-3} $ \\
		~ & $\Delta_\text{AIM}$ &  $5.68\times 10^{-6} $ & $ 6.30\times 10^{-6} $ & $ 7.53\times 10^{-6} $ & $ 1.05\times 10^{-5} $ & $ 1.41\times 10^{-4}  $ \\
        ~ & $\Delta_\text{WA}$ &  $ 0.016220\%$ & $ 0.001073\% $ & $ 0.026520\%$ & $ 0.046700\%$ & $ 0.050870\% $ \\
		\hline
		\multirow{2}*{$ 2 M \omega_{03} $} & $\Delta_\text{WKB}$ & $ 1.20\times 10^{-5} $ & $ 1.03 \times 10^{-5}$ & $ 4.26\times 10^{-5} $ & $ 1.02\times 10^{-4} $ & $ 2.00 \times 10^{-4}$ \\
		~ & $\Delta_\text{AIM}$ &  $ 8.91\times 10^{-6} $ & $ 1.05\times 10^{-6} $ & $1.40\times 10^{-6}  $ & $2.37\times 10^{-6}  $ & $1.86\times 10^{-5}   $ \\
        ~ & $\Delta_\text{WA}$ &  $ 0.000166\%$ & $ 0.000134\% $ & $ 0.000144\%$ & $ 0.000194\%$ & $ 0.000393\% $ \\
		\hline
        \multirow{2}*{$ 2 M \omega_{04} $} & $\Delta_\text{WKB}$ & $ 1.21\times10^{-6} $ & $ 2.10\times10^{-6} $ & $ 5.04\times10^{-6} $ & $ 5.21\times10^{-6} $ & $  1.69\times10^{-5}$\\
		~ & $\Delta_\text{AIM}$ &  $ 2.51\times 10^{-7} $ & $ 3.17\times 10^{-7}  $ & $ 4.65\times 10^{-7} $ & $ 9.21\times 10^{-7} $ & $ 4.67\times 10^{-6}  $ \\
        ~ & $\Delta_\text{WA}$ &  $ 0.000029\%$ & $ 0.000028\% $ & $0.000320\% $ & $ 0.000041\%$ & $ 0.000135\% $ \\
		\hline
	\end{tabular}
	\caption{The numerical errors of the quasinormal frequencies obtained from the WKB approximation method and the asymptotic iteration method, respectively. And the relative errors between quasinormal frequencies obtained from the two methods.}
	\label{T-errs}
\end{table*}

To study the axial gravitational perturbation of the holonomy-corrected Schwarzschild black hole more comprehensively, we investigate the numerical evolution of a Gaussian wave packet on the background spacetime under the axial gravitational perturbation. In the time domain, we can rewrite the Schr\"{o}dinger-like master equation \eqref{master-eq} as
\bqn
\frac{\p^2 \Psi}{\p r^2_{\ast}} - \f{\p^2 \Psi}{\p t^2} - V (r_\ast) \Psi =0.
\eqn
Under the light-cone coordinates $u = t - r_{\ast}$ and $ v = t + r_{\ast}$ \cite{Gundlach:1993tp}, the above equation becomes
\bqn\lb{ringdown-eq}
4 \frac{\p ^2 \Psi (u, v)}{\p u \p v} - V (u, v) \Psi (u,v) = 0.
\eqn
We assume that the wave packet is a Gaussian pulse centered in $v_c$ and having width $\beta$:
\bqn
\Psi (u, 0) = 0~~\text{and}~~\Psi (0,v) = \tx{exp} \left( - \frac{(v-v_{c})^2}{2 \beta^2} \right).
\eqn
Then, we set $v_c/2M = 10,~\beta/2M = 1,~ M=1/2,~l=2$, choose the observer located at $r = 10 r_\text{H}$, and numerically solve the partial differential equation (\ref{ringdown-eq}) to generate the ringdown waveform with different values of the parameter $r_0$.~We employ a finite difference method to discretize Eq. (\ref{ringdown-eq})  and utilize a time evolution scheme to track the wave packet over time. Here, we apply the discretization  of Gundlach-Price-Pullin \cite{Gundlach:1993tp}
\bqn
\Psi (u + \Delta, v+ \Delta) &=& \Psi (u, v+ \Delta) + \Psi (u + \Delta, v) - \Psi (u, v) \nb\\
&&- \Delta^2 V(u,v) \nb\\
&&\times\f{\Psi (u + \Delta, v) + \Psi (u + \Delta, v)}{4} 
\eqn
to integrate Eq.~(\ref{ringdown-eq}), where $\Delta$ is the step size  during the numerical procedure. This method ensures that the time derivatives are centered in time and maintain stability \cite{Gundlach:1993tp}.

\begin{figure}[h]\centering
	\includegraphics[scale =0.28]{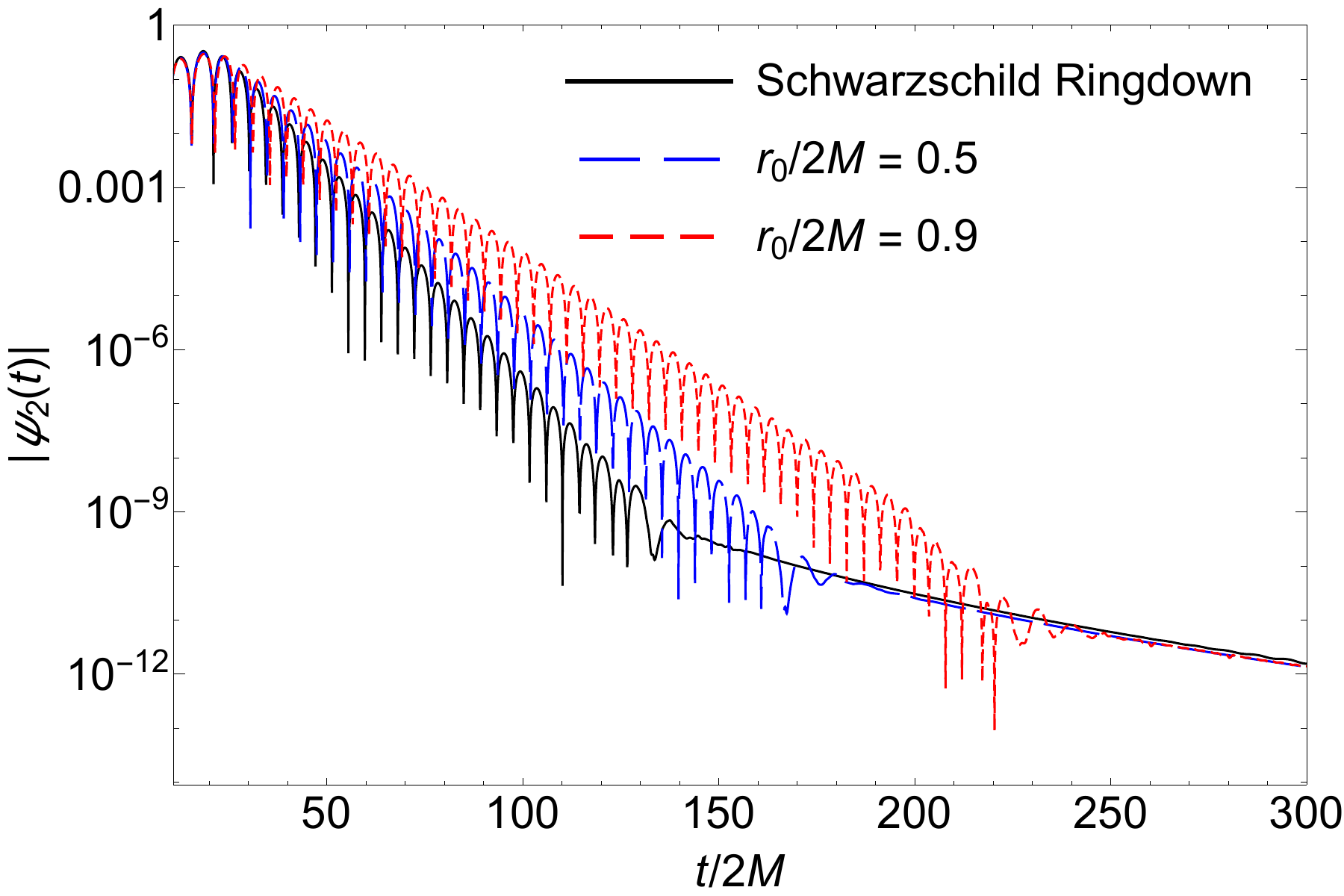}
	\caption{The time evolution of the wave function $\Psi (t) (l=2)$ of the axial gravitational perturbation of the holonomy-corrected Schwarzschild black hole with different values of the parameter $r_0$, evaluated at $r = 10 r_\text{H}$. The black curve ($r_0 = 0$) shows the Schwarzschild ringdown case.}
	\label{ringdown-plot}
\end{figure}

The evolution of the Gaussian pulse is shown in Fig.~\ref{ringdown-plot}. One can see that the decay of the wave packet becomes slower as the parameter $r_0$ increases, which agrees with the fact that $r_0$ negatively affects the imaginary part of the quasinormal frequencies. The fundamental mode $\omega_{02}$ mainly controls the behavior of the ringdown waveforms. Without loss of generality, we use a modified exponentially decaying function $ e^{\omega_I t} A \sin(\omega_R + B)$ to fit the data in Fig.~\ref{ringdown-plot} and obtain the value of $\omega_{02}$ with different values of the parameter $r_0$. The results are shown in Tab.~\ref{ringdown}. One can find that the fitting values of the fundamental mode $\omega_{02}$  with different values of the parameter $r_0$ in Tab.~\ref{ringdown} agree well with the results obtained by using the WKB approximation method and the asymptotic iteration method. Finally, as shown in Fig.~\ref{ringdown-plot}, the late-time behavior of the evolution of the Gaussian pulse is a power-law tail. For the case of the Schwarzschild black hole, the power-law relation is $\Psi = t^{-(2l+3)}$ \cite{Ching:1995tj}. Here, we use a modified power-law function 
$t^{-(2l + \zeta)}$ to fit the late time data in Fig.~\ref{ringdown-plot} and obtain the value of $\zeta$ with different values of the parameter $r_0$. The results are shown in Tab.~\ref{tail}. It shows that the parameter $r_0$ does not affect the late-time behavior of the evolution of the wave packet under the axial gravitational perturbation of the holonomy-corrected Schwarzschild black hole. This agrees with the conclusion for the massless scalar field perturbation of the holonomy-corrected Schwarzschild black hole \cite{Moreira:2023cxy}.

\begin{table*}[t]
	\renewcommand\arraystretch{1.5} 
	\centering
	\begin{tabular}{|c|c|c|c|c|}
		\hline
		\multicolumn{2}{|c|}{$r_0 / 2M$} & $0$  & $0.5$  & $0.9$  \\
		\hline
		\multirow{3}*{$ 2 M \omega_{02} $}   & WKB & $ 0.747239 - 0.177782 i$ & $ 0.748530 - 0.144919 i$ & $ 0.747798 - 0.111396 i$ \\
		~& AIM & $ 0.747343 - 0.177928 i$ & $ 0.748435 - 0.144864 i$ & $ 0.747847 - 0.111449 i$ \\
		~& Fitting & $ 0.747216 - 0.178764 i$ & $ 0.748509 - 0.145371 i$ & $ 0.747844 - 0.111992 i$ \\
		\hline
	\end{tabular}
	\caption{The quasinormal frequencies of fundamental modes $\omega_{02}$ with different values of $r_0$ calculated by fitting the data in Fig.~\ref{ringdown-plot}, using the WKB approximation method, and using the asymptotic iteration method.}
	\label{ringdown}
\end{table*}

\begin{table*}[t]
	\renewcommand\arraystretch{1.5} 
	\centering
	\begin{tabular}{|c|c|c|c|}
		\hline
		$r_0 /2M$ & $0$  & $0.5$  & $0.9$  \\
		\hline
		$ \zeta $ & $ 3.03098 $ & $ 3.05766 $ & $ 3.02208 $ \\
		\hline
	\end{tabular}
	\caption{The power-law parameter $\zeta$ with different values of $r_0$ calculated by fitting the late-time data in Fig.~\ref{ringdown-plot}.}
	\label{tail}
\end{table*}

\section{Parameterized quasinormal frequencies}\lb{sec3}
\renewcommand{\theequation}{3.\arabic{equation}}
\setcounter{equation}{0} 

Assuming a background spacetime has spherical symmetry and differs slightly from the Schwarzschild black hole, Cardoso $et~al.$ proposed the parameterized framework to calculate the quasinormal frequencies of the black hole \cite{Cardoso:2019mqo, McManus:2019ulj}. This framework provides a new approach to studying the QNMs of spherically symmetric black holes. In this section, we obtain the parameterized approximation of the quasinormal frequencies of the axial gravitational perturbation of the holonomy-corrected Schwarzschild black hole and verify its effectiveness.

Defining
\bqn
F(r) = \f{dr}{d r_\ast} = f(r) \sqrt{g(r)},
\eqn 
one can rewrite the master equation (\ref{master-eq}) of the axial gravitational perturbation of the holonomy-corrected Schwarzschild black hole as
\bqn\lb{master-eq-2}
F \f{d}{d r} \left( F \f{d \Psi}{dr} \right) + \left( \omega^2 - F \bar{V} \right) \Psi = 0,
\eqn 
where $\bar{V}(r) \equiv V(r) /F(r)$. Considering the holonomy-corrected Schwarzschild black hole must differ from the Schwarzschild black hole slightly, one can assume
\bq\lb{F(r)}
F(r) = \left( 1- \f{r_H}{r} \right) \left[ 1 + \epsilon(r) \right],
\eq 
where
\bqn
\epsilon(r) = \sqrt{g(r)} - 1
\eqn
is a small quantity. And one can write $\bar{V}$ as 
\bq\lb{V-expand}
\bar{V} = V_\text{GR} + \delta \bar{V},
\eq
where
\bqn\lb{potential-GR}
V_{\text{GR}} &=& \f{l(l+1)}{r^2} - \f{3 r_H}{r^3}
\eqn 
is the effective potential for the axial gravitational perturbation of the Schwarzschild black hole in general relativity, and $\delta \bar{V}$ is a small correction.
Defining
\bq
\phi = \left[ 1 + \epsilon (r) \right]^{1/2} \Psi,
\eq
with Eq. (\ref{F(r)}), one can write Eq. (\ref{master-eq-2}) as
\bq\lb{master-eq-3} 
f \f{d}{d r} \left( f \f{d \phi}{d r} \right) + \left[ \f{\omega^2}{( 1 + \epsilon )^2} - f \mathcal{V} \right] \phi = 0, 
\eq
where
\bqn\lb{potential2}
\mathcal{V} = \f{\bar{V}}{ 1 + \epsilon} - \f{f  \epsilon '^2 - 2 (1 + \epsilon) (f \epsilon' )'}{ 4 ( 1 + \epsilon )^2}.
\eqn
The frequency-dependent term in Eq. \eqref{master-eq-3} could be expanded at $r=r_H$ as 
\bq\lb{qnm2}
\f{\omega^2}{( 1 + \epsilon )^2} = \omega^2 [1 - 2 \epsilon(r_\text{H})] - 2 \omega^2 [\epsilon(r) - \epsilon(r_H)],
\eq
where the high-order terms are ignored. The first term on the right-hand side in Eq. \eqref{qnm2} is a rescaling of the quasinormal frequencies. And one can absorb the second term on the right-hand side in Eq. \eqref{qnm2} in the potential \eqref{potential2}. Then, one can rewrite Eq. (\ref{master-eq-3}) as
\bq\lb{master-eq-4} 
f \f{d}{d r} \left( f \f{d \phi}{d r} \right) + \left\{ \omega^2 [ 1 + \epsilon(r_\text{H}) ]^2 - f \mathcal{V}_\text{new} \right\} \phi = 0. 
\eq
With Eq. \eqref{V-expand}, to leading order, the new potential in Eq. \eqref{master-eq-4} is
\bqn\lb{new-potential}
\mathcal{V}_\text{new} &=& V_\text{GR} + \delta \bar{V} - V_\text{GR} \epsilon(r) + \f{1}{2} \left[ f(r) \epsilon(r) '\right]' \nb\\
&&+ \frac{2 \omega_0^2 [ \epsilon(r) - \epsilon(r_H)]}{f(r)},
\eqn
where we set the frequency $\omega$ in the $\omega$-dependent term equal to $\omega_0$ (the frequencies for the case of the axial gravitational perturbation of the Schwarzschild black hole in general relativity).
This new potential \eqref {new-potential} could be regarded as the corresponding effective potential \eqref{potential-GR} of the Schwarzschild black hole in general relativity with a correction
\bq
\mathcal{V}_\text{new} = V_\text{GR} + \delta \mathcal{V},
\eq
where the correction term is
\bq\lb{perturbing potential}
\delta \mathcal{V} = \delta \bar{V} - V_{\text{GR}} \epsilon(r) + \f{1}{2} \left[ f(r) \epsilon(r) '\right]' + \frac{2 \omega_0^2 [ \epsilon(r) - \epsilon(r_H)]}{f(r)}.
\eq 

In the parameterized framework of QNMs \cite{Cardoso:2019mqo}, one can write the correction term (\ref{perturbing potential}) as the power-law form  
\bqn\lb{perturbing potential-power-law}
\delta \mathcal{V} = \f{1}{r^2_H} \sum ^\infty_{j=0} \alpha^{-}_j \left( \f{r_H}{r} \right)^j,
\eqn 
where $\alpha^{-}_j$ are constant coefficients. In this work, without loss of generality, we only keep the terms with $0 \leq j \leq 5$ in Eq. (\ref{perturbing potential-power-law}), and we obtain the corresponding coefficients as
\bqn
\alpha^{-}_0 &=& -2 r_H^2 \epsilon(r_H) \omega_0^2, \nb\\
\alpha^{-}_1 &=& -r_H [ r_0 + 2 r_H \epsilon(r_H) ] \omega_0^2, \nb\\
\alpha^{-}_2 &=& - \f{1}{4} [(r_0^2 + 4 r_0 r_H + 8 r_H^2 \epsilon(r_H)] \omega_0^2 , \nb\\
\alpha^{-}_3 &=&  \f{1}{8 r_H} [  8 r_0 (l + l^2 -3  - r_H^2 \omega_0^2) -r_0^3 \omega_0^2 - 2 r_0^2 r_H \omega_0^2 \nb\\
&&- 16 r_H^3 \epsilon(r_H) \omega_0^2 ], \nb\\
\alpha^{-}_4 &=&\f{1}{ 64 r_H^2} [ 8 r_0^2 (4 l + 4 l^2 -13  - 2 r_H^2 \omega_0^2) + r_0 (80 r_H \nb\\
&&- 64 r_H^3 \omega_0^2) -5 r_0^4 \omega_0^2 - 8 r_0^3 r_H \omega_0^2 - 128 r_H^4 \epsilon(r_H) \omega_0^2 ] , \nb\\
\alpha^{-}_5 &=& \frac{1}{128 r_H^3} [ 8 r_0^3 (6 l + 6 l^2 -21  - 2 r_H^2 \omega_0^2) \nb\\
&&- 32 r_0^2 r_H ( r_H^2 \omega_0^2 -3) - 7 r_0^5 \omega_0^2 - 10 r_0^4 r_H \omega_0^2  \nb\\
&&- 128 r_0 r_H^4 \omega_0^2 + 256 r_H^5 \epsilon (r_H) \omega_0^2 ]. \nb\\
\eqn 
Then, the parameterized approximation of the quasinormal frequencies of the axial gravitational perturbation of the holonomy-corrected
Schwarzschild black hole is 
\bqn\lb{QNM-expand}
\omega_{\text{p}} &=& [1 + \epsilon(r_H) ] \left( \omega_0 + \sum ^\infty_{j=0} \alpha^{-}_j e^{-}_j \right) \nb\\
&\simeq& \sqrt{g(r_H)} \left( \omega_0 + \sum ^{5}_{j=0} \alpha^{-}_j e^{-}_j \right),
\eqn 
where $e^{-}_j$ are the complex basis. The values of $e^{-}_j$ used in this work are shown in Tab. \ref{basis} (The full set of basis is provided online by Cardoso and Berti \cite{basis}). To see the corrections to the quasinormal frequencies from different terms in Eq. (\ref{QNM-expand}), we calculate the percentages of different order terms and $\omega_0^{-}$. We list the numerical results in Tab. \ref{The percentages of different terms}. One can find that: i) No corrections to $\omega_0^{-}$ when $r_0 = 0$. ii) The $\alpha^{-}_0 e^{-}_0$ and $\alpha^{-}_3 e^{-}_3$ play the dominant roles when $r_0 \neq 0$. iii) The corrections to $\omega_0^{-}$ from different terms are all proportionate to the order of magnitude of the parameter $r_0$. iv) The corrections to $\omega_0^{-}$ from different terms are small when $r_0 \leq 10^{-2}$.   

\begin{table*}[t]
\renewcommand\arraystretch{1.5} 
	\centering
	\begin{tabular}{|c|c|c|c|}%{|c{1.0cm}|c{3.0cm}|c{2.0cm}|c{3.0cm}|c{2.0cm}|}
		\hline
        $ j $ &$  r_H e^{-}_ j (l=2) $ & $  r_H e^{-}_ j (l=3) $ & $ r_H e^{-}_ j (l=4)$   \\
		\hline
        $ 0 $ &  $ 0.24725196828088 + 0.09264307282584 i $  & $ 0.14442743026294 + 0.03677032261872 i $ & $ 0.1050910884042 + 0.0202958065346i $   \\
        \hline
        $ 1 $ &  $ 0.15985477262517 + 0.01820847613128 i $  & $ 0.09576831949608 + 0.00860354722266 i $ & $ 0.0699556694195 + 0.0050316213505 i $   \\
        \hline
        $ 2 $ &  $ 0.09663223435342 - 0.00241549603538 i $  & $ 0.06147249927577 - 0.00061952324973i $ & $ 0.0456954400020 - 0.0002146177462 i $   \\
        \hline
        $ 3 $ &  $ 0.05849078435691 - 0.00371786167568 i $  & $ 0.03929284560364 - 0.00202787827974 i $ & $ 0.0297480560127 - 0.0011736685414 i $   \\
        \hline
        $ 4 $ &  $ 0.03667943748105 - 0.00043869803165 i $  & $ 0.02543229547295 - 0.00096084572194 i $ & $ 0.0194908220414 - 0.0006562131653 i $   \\
        \hline
        $ 5 $ &  $ 0.02403794871363 + 0.00273079210136 i $  & $ 0.01678541589555 + 0.00045258539650 i $ & $ 0.0129120394104 + 0.0001390032920 i $ \\
        \hline
	\end{tabular}
    \caption{The basis $e^{-}_j$ of the axial gravitational perturbation with $0 \leq j \leq 5$ and $l=2,~3,~4$ \cite{basis}.}
    \lb{basis}
\end{table*}

\begin{table*}[t]
\renewcommand\arraystretch{1.5} 
	\centering
	\begin{tabular}{|c|c|c|c|c|c|c|}%{|c{1.0cm}|c{3.0cm}|c{2.0cm}|c{3.0cm}|c{2.0cm}|}
		\hline
        $ r_0 $ & $ | \alpha_{0}^{-} e_{0}^{-} / \omega_{0}^{-}  | $ & $ | \alpha_{1}^{-} e_{1}^{-} / \omega_{0}^{-}  | $ & $ | \alpha_{2}^{-} e_{2}^{-} / \omega_{0}^{-}  | $ & $ | \alpha_{3}^{-} e_{3}^{-} / \omega_{0}^{-}  | $& $ | \alpha_{4}^{-} e_{4}^{-} / \omega_{0}^{-}  | $ & $ | \alpha_{5}^{-} e_{5}^{-} / \omega_{0}^{-}  | $ \\
		\hline
        $0$ & $ 0 $  &  $ 0 $ &  $ 0 $  & $ 0 $ &  $ 0 $ & $ 0 $  \\
        \hline
         $ 10^{-5} $ & $ 2 \times 10^{-6} $  &  $ 3 \times 10^{-12} $ &  $ 5 \times 10^{-18} $  & $ 2 \times 10^{-6} $ &  $ 6 \times 10^{-7} $ & $ 2 \times 10^{-12} $  \\
        \hline
         $ 10^{-4} $ & $ 0.002 \% $  &  $ 3 \times 10^{-10} $ &  $ 9 \times 10^{-15} $  & $ 0.005 \% $ &  $ 6 \times 10^{-6} $ & $ 2 \times 10^{-10} $  \\
        \hline
         $ 10^{-3} $ & $ 0.02 \% $  &  $ 3 \times 10^{-8} $ & $9 \times 10^{-12} $ &  $ 0.02 \% $  & $ 0.006 \% $ &  $ 2 \times 10^{-8} $  \\
        \hline
         $ 10^{-2} $ & $ 0.20 \% $  &  $ 3 \times 10^{-6} $ &  $ 9 \times 10^{-9} $  & $ 0.23 \% $ &  $ 0.06 \% $ & $ 2 \times 10^{-6} $  \\
        \hline
         $ 0.1 $ & $ 2.08 \% $  &  $ 0.03 \% $ &  $ 9.9 \times 10^{-6} $  & $ 2.29 \% $ &  $ 0.66 \%  $ & $ 0.03 \%  $  \\
        \hline
         $ 0.5 $ & $ 11.88 \% $  &  $ 1.06 \% $ &  $ 0.17 \% $  & $ 11.47 \% $ &  $ 4.63 \% $ & $ 0.96\% $  \\
        \hline
	\end{tabular}
    \caption{The percentages of different order terms and $\omega_0^{-}$, with the parameter $r_0$.}
    \lb{The percentages of different terms}
\end{table*}

Then, one can quantify the accuracy of the parameterized approximation by defining the relative errors of the real parts and the imaginary parts of quasinormal frequencies calculated by the asymptotic iteration method and the parameterized approximation (\ref{QNM-expand}) as
\bqn
\Delta_{\text{Re}} &=& \left| \f{\text{Re}(\omega_\text{p})}{\text{Re}(\omega_\text{AIM})} - 1 \right|, \lb{accuracy-1}\\
\Delta_{\text{Im}} &=& \left| \f{\text{Im}(\omega_\text{p})}{\text{Im}(\omega_\text{AIM})} - 1 \right|.  \lb{accuracy-2}
\eqn 
With Eqs. \eqref{accuracy-1} and \eqref{accuracy-2}, we obtain the relative errors of the real and imaginary parts of $\omega_{02}$, $\omega_{03}$, and $\omega_{04}$ calculated by the asymptotic iteration method and the parameterized approximation with 30th-order expansion with different values of the parameter $r_0$. The numerical results are listed in Tab. \ref{The relative errors}. 
\begin{table*}[t]
\renewcommand\arraystretch{1.5} 
	\centering
	\begin{tabular}{|c|c|c|c|c|c|c|}%{|c{1.0cm}|c{3.0cm}|c{2.0cm}|c{3.0cm}|c{2.0cm}|}
		\hline
        $ r_0 $ & $ \Delta_{\text{Re}} (\omega_{02}) $ & $ \Delta_{\text{Im}} (\omega_{02}) $ & $ \Delta_{\text{Re}} (\omega_{03})$ & $ \Delta_{\text{Im}}(\omega_{03}) $& $ \Delta_{\text{Re}}(\omega_{04}) $ & $ \Delta_{\text{Im}}(\omega_{04}) $ \\
		\hline
        $0$ & $ 0 $  &  $ 0 $ &  $ 0 $  & $ 0 $ &  $ 0 $ & $0$  \\
        \hline
         $ 10^{-5} $ &  $ 0.0005 \% $ &  $0.0005 \% $  & $ 0.0005 \% $ &  $ 0.0001 \% $  & $ 0.0005 \% $ &  $ 0.0002 \% $  \\
        \hline
         $ 10^{-4} $ &  $ 0.005 \% $ &  $0.005 \% $  & $ 0.005 \% $ &  $ 0.001 \% $  & $ 0.005 \% $ &  $ 0.002 \% $  \\
        \hline
         $ 10^{-3} $ & $ 0.05 \% $ &  $ 0.05 \% $  & $ 0.05 \% $ &  $ 0.01 \% $ &   $ 0.05 \% $ &  $ 0.02 \% $   \\
        \hline
         $ 10^{-2} $ & $ 0.50 \% $ &  $ 0.50 \% $  & $ 0.47 \% $ &  $ 0.14 \% $ &  $ 0.48 \% $ &  $ 0.16 \% $    \\
        \hline
         $ 0.1 $ & $ 5.34 \% $ &  $ 5.23 \% $  & $ 5.06 \% $ &  $ 1.69 \% $  & $ 5.10 \% $ &  $ 1.90 \% $   \\
        \hline
         $ 0.5 $ & $ 34.68 \% $ &  $ 32.13 \% $  & $ 33.68 \% $ &  $ 18.35 \% $  & $ 33.77 \% $ &  $ 19.06 \% $   \\
        \hline
	\end{tabular}
    \caption{The relative errors of the real and imaginary parts of $\omega_{02}$, $\omega_{03}$, and $\omega_{04}$ obtained from the parameterized approximation compared with that from the asymptotic iteration method with 30-th order expansions, with different values of the parameter $r_0$.}
    \lb{The relative errors}
\end{table*}
One can find that: i) All of $\Delta_{\text{Re}}$ and $\Delta_{\text{Im}}$ vanish when $r_0 = 0$. ii) For the same value of the parameter $r_0$, $\Delta_{\text{Re}}$ and $\Delta_{\text{Im}}$ for $\omega_{02}$, $\omega_{03}$, and $\omega_{04}$ are of the same order. iii) Both $\Delta_{\text{Re}}$ and $\Delta_{\text{Im}}$ are proportional to the order of magnitude of the parameter $r_0$. iv) The quasinormal frequencies $\omega_{02}$, $\omega_{03}$, and $\omega_{04}$ obtained from the asymptotic iteration method and the parameterized approximation agree well with each other when $r_0 \leq 10^{-2}$. Finally, we conclude that $r_0 \leq 10^{-2}$ is a necessary condition for the parameterized approximation (\ref{QNM-expand}) of the quasinormal frequencies of the axial gravitational perturbation of the holonomy-corrected Schwarzschild black hole to be valid. We also calculate the relative errors of the real and imaginary parts of $\omega_{02}$, $\omega_{03}$, and $\omega_{04}$ obtained from the parameterized approximation compared with that from the asymptotic iteration method with different order expansions, with $r_0 = 0.1,~10^{-2},~ \text{and}~ 10^{-3}$. The results are listed in Tab.~\ref{errors-2}. It shows that the numerical errors will not affect the conclusion.

\begin{table*}[t] \renewcommand\arraystretch{1.5}
	\centering
	\begin{tabular}{|c|c|c|c|c|c|c|c|}%{|c{1.0cm}|c{3.0cm}|c{2.0cm}|c{3.0cm}|c{2.0cm}|}
		\hline
		\multicolumn{2}{|c|}{$ r_0 $} & $ \Delta_{\text{Re}} (\omega_{02}) $ & $ \Delta_{\text{Im}} (\omega_{02}) $ & $ \Delta_{\text{Re}} (\omega_{03})$ & $ \Delta_{\text{Im}}(\omega_{03}) $& $ \Delta_{\text{Re}}(\omega_{04}) $ & $ \Delta_{\text{Im}}(\omega_{04}) $ \\
		\hline
		\multirow{3}*{$ 10^{-3} $} & 25-th order & $ 0.048965 \% $ &  $ 0.049480 \% $  & $ 0.046988 \% $ &  $ 0.013334 \% $ &   $ 0.047539 \% $ &  $ 0.015553 \% $   \\
		~ & 30-th order & $ 0.050033 \% $ &  $ 0.050021 \% $  & $ 0.047095 \% $ &  $ 0.013393 \% $ &   $ 0.047558 \% $ &  $ 0.015620 \% $   \\
		~ & 35-th order & $ 0.050270 \% $ &  $ 0.048724 \% $  & $ 0.047107 \% $ &  $ 0.013234 \% $ &   $ 0.047561 \% $ &  $ 0.015592 \% $   \\
		\hline
		\multirow{3}*{$ 10^{-2} $} & 25-th order & $ 0.502225 \% $ &  $ 0.501604 \% $  & $ 0.473954 \% $ &  $ 0.136886 \% $ &  $ 0.478626 \% $ &  $ 0.159014 \% $    \\
		~ & 30-th order & $ 0.503288 \% $ &  $ 0.502174 \% $  & $ 0.474060 \% $ &  $ 0.136949 \% $ &  $ 0.478645 \% $ &  $ 0.159082 \% $    \\
		~ & 35-th order & $ 0.503526 \% $ &  $ 0.500887 \% $  & $ 0.474073 \% $ &  $ 0.136790 \% $ &  $ 0.478648 \% $ &  $ 0.159055 \% $    \\
		\hline
		\multirow{3}*{$ 0.1 $} & 25-th order  & $ 5.336820 \% $ &  $ 5.226400 \% $  & $ 5.060120 \% $ &  $ 1.689570 \% $  & $ 5.101060 \% $ &  $ 1.897400 \% $   \\
		~ & 30-th order & $ 5.337830 \% $ &  $ 5.227270 \% $  & $ 5.060220 \% $ &  $ 1.689680 \% $  & $ 5.101080 \% $ &  $ 1.897480 \% $   \\
		~ & 35-th order & $ 5.338070 \% $ &  $ 5.226090 \% $  & $ 5.060230 \% $ &  $ 1.689520 \% $  & $ 5.101080 \% $ &  $ 1.897460 \% $   \\
		\hline
	\end{tabular}
	\caption{The relative errors of the real and imaginary parts of $\omega_{02}$, $\omega_{03}$, and $\omega_{04}$ obtained from the parameterized approximation compared with that from the asymptotic iteration method with different order expansions, with different values of the parameter $r_0$.}
	\lb{errors-2}
\end{table*}

\section{Greybody factor and  Hawking radiation}\lb{sec4}
\renewcommand{\theequation}{4.\arabic{equation}}
\setcounter{equation}{0} 

The greybody factor is the probability for an outgoing wave to reach an observer at infinity or the probability for an incoming wave to be absorbed by the black hole \cite{Konoplya:2019ppy, Konoplya:2019hlu, Konoplya:2023moy}. Here, we define the boundary conditions for the waves scattered by the effective potential \eqref{V} as
\bqn
\Psi (r_\ast) &=& T(\omega) e^{- i \omega r_\ast},~~~~~~~~ ~~~~~~ r_\ast \rightarrow - \infty, \\
\Psi (r_\ast) &=& e^{- i \omega r_\ast} + R(\omega) e^{i \omega r_\ast},~~~~  r_\ast \rightarrow \infty,
\eqn
where $R(\omega)$ and $T(\omega)$ are the reflection and transmission coefficients, respectively. And they should satisfy $|R(\omega)|^2 + |T(\omega)|^2 =1$ since the conservation of probability. With the WKB approximation
approach, one can obtain 
\bqn
|R(\omega)|^2 &=& \f{1}{1 + e^{-2 i \pi \mathcal{K}}}, \\
|T(\omega)|^2 &=& \f{1}{1 + e^{2 i \pi \mathcal{K}}}.
\eqn
The parameter $\mathcal{K}$ is determined by \cite{Konoplya:2019hlu}
\bqn
\mathcal{K} = \f{i (\omega^2 - V_0)}{\sqrt{-2 V_0^{''}}} - \sum_{j} \Lambda_j,
\eqn
where $V_0$ is the maximum of the effective potential, $V_0^{''}$ is the second derivative of the effective potential in its maximum
to the tortoise coordinate $r_\ast$, and $\Lambda_j$ are the higher order WKB correction terms. The greybody factor is defined as
\bqn
|A|^2 = |T(\omega)|^2.
\eqn

\begin{figure}[!t]
	\centering
	\subfigure[]{\includegraphics[scale =0.28]{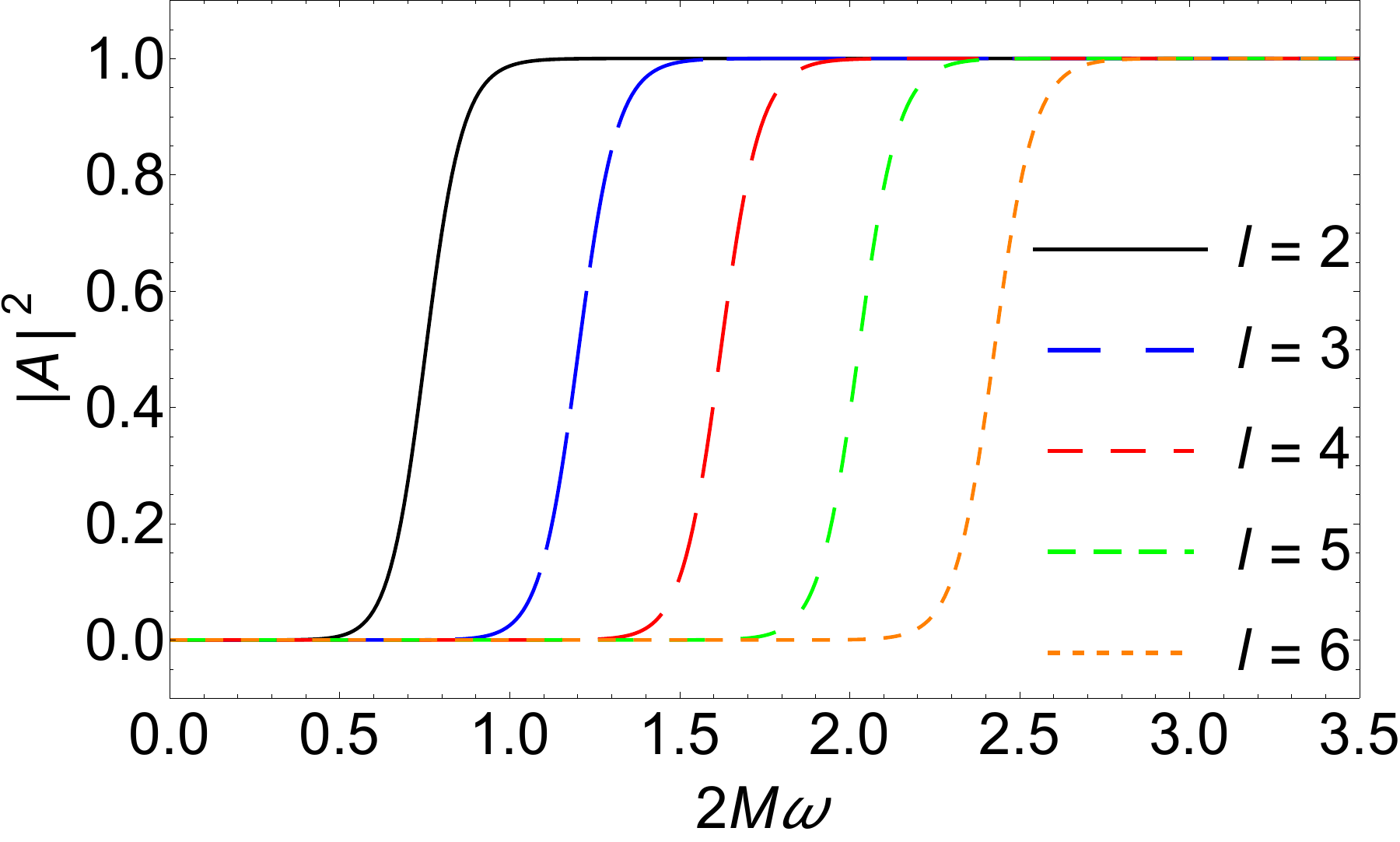}}
	\subfigure[]{\includegraphics[scale =0.28]{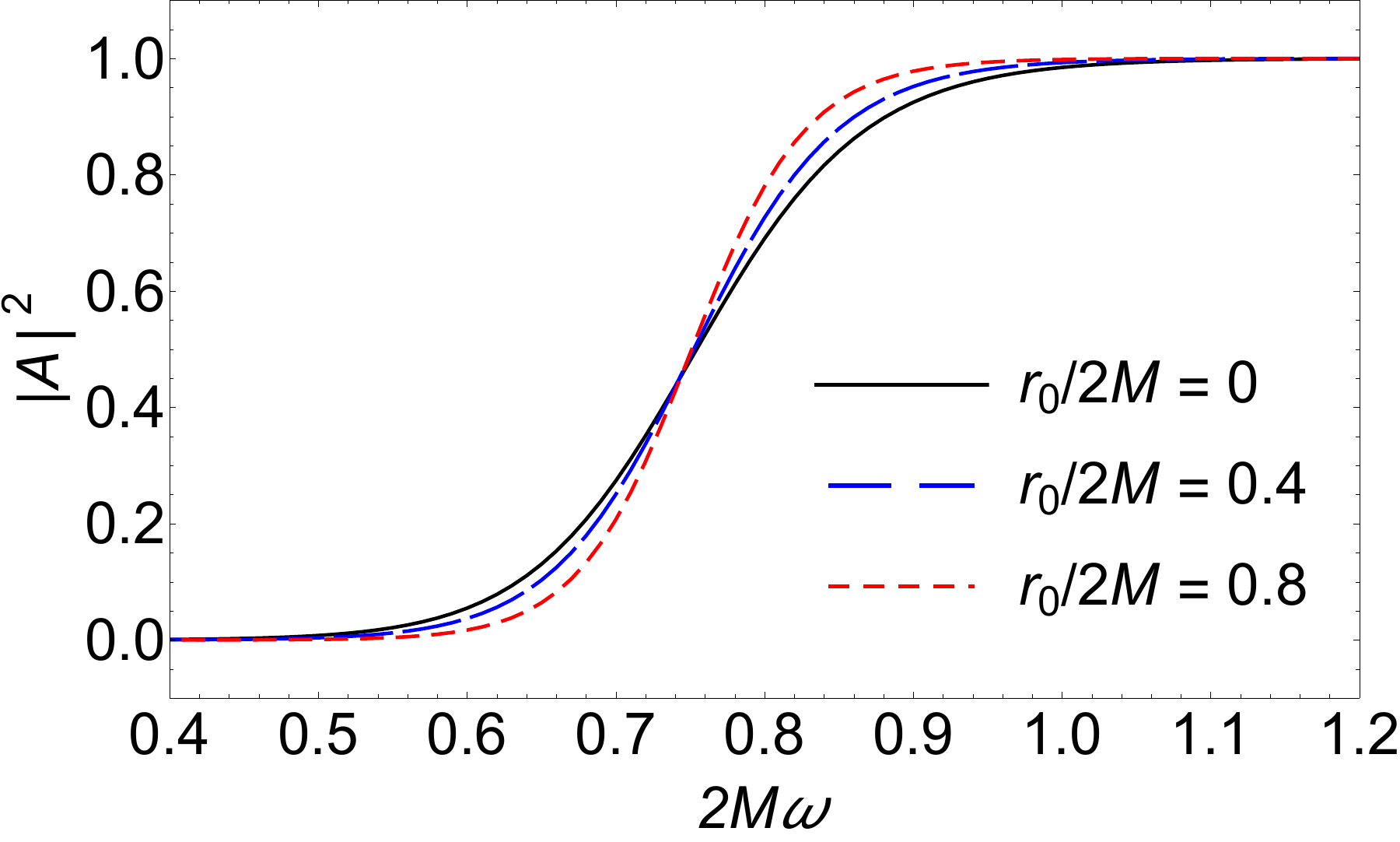}}
	\caption{(a) The greybody factor versus the frequency $\omega$ for different values of $l$, with $M = 1/2$ and $r_0 = 0.1$. (b) The greybody factor versus the frequency $\omega$ for different values of $r_0$, with $M = 1/2$ and $l = 2$.}
	\label{greybody factor}
\end{figure}

We calculate the greybody factor for the effective potential \eqref{V} versus frequency $\omega$, with different values of $l$ and the parameter $r_0$. The numerical results are shown in Fig. \ref{greybody factor}. From Fig. \ref{greybody factor} (a), one can see that the larger the value of $l$, the greater the frequency value corresponding to the non-zero greybody factor. And from Fig. \ref{greybody factor} (b), one can see that when $r_0$ is large, the greybody factor is small for small $\omega$, but it is large for large $\omega$.

\begin{figure}[!t]
	\centering
	\includegraphics[scale =0.28]{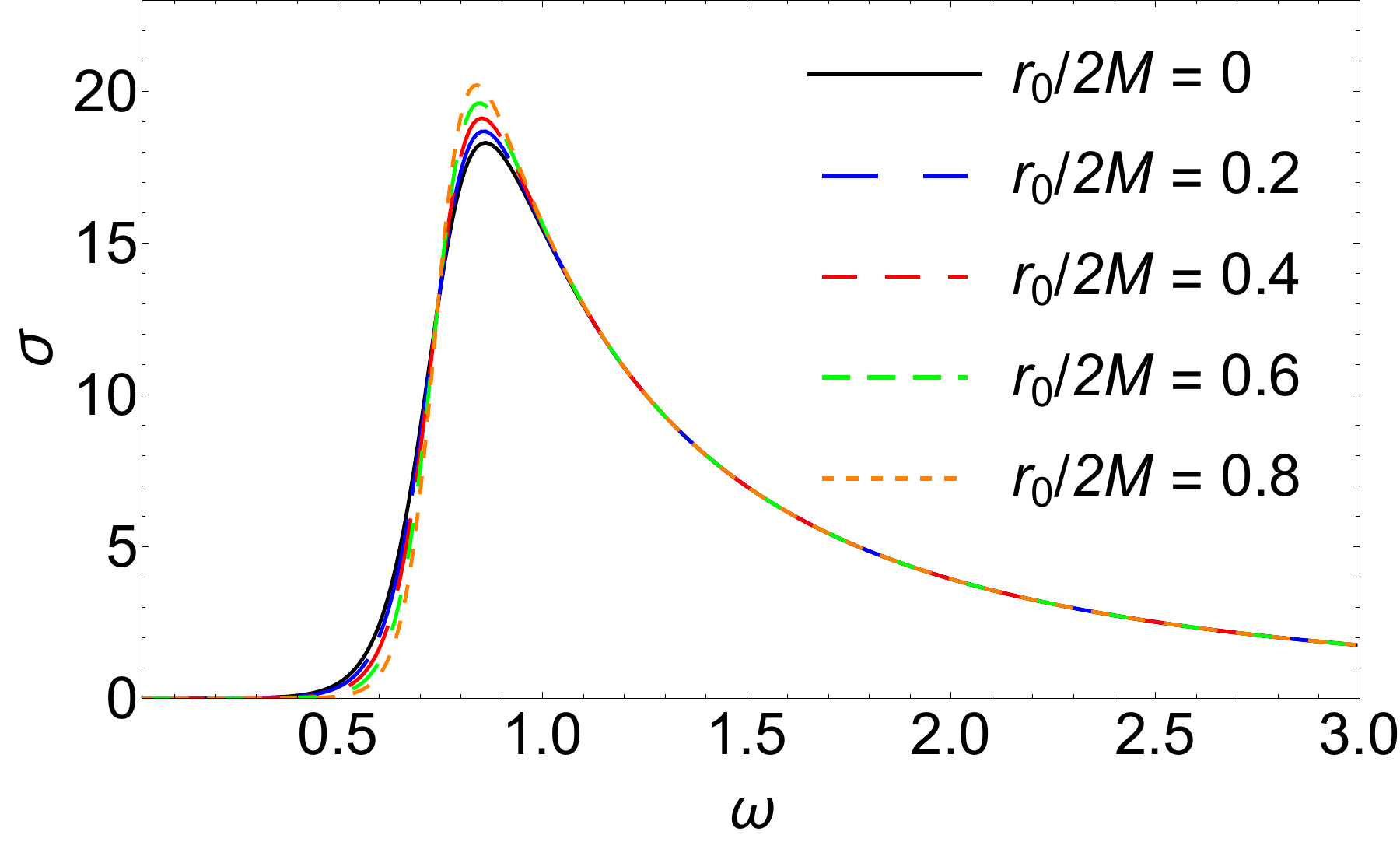}
    \includegraphics[scale =0.28]{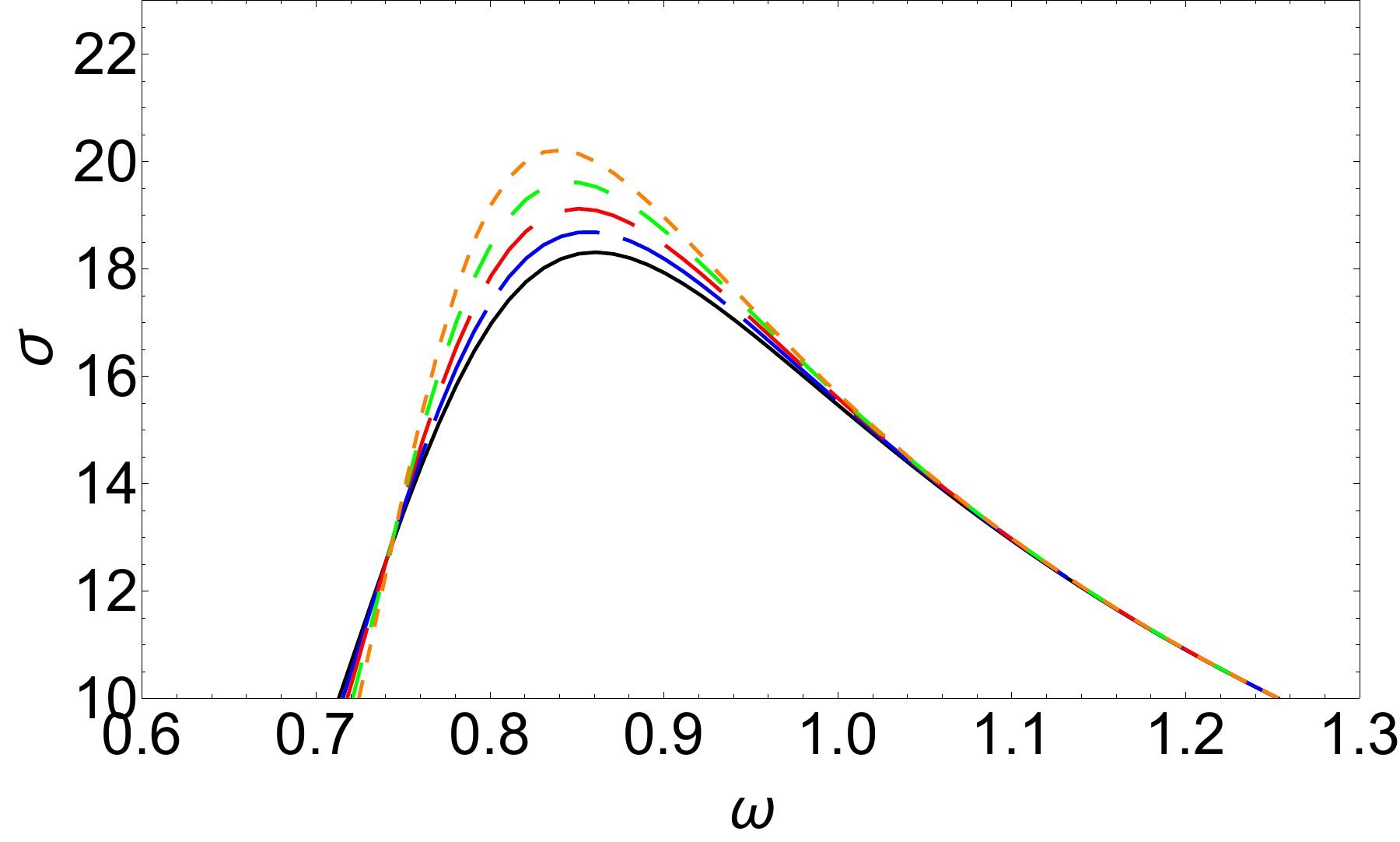}
	\caption{The partial absorption cross section as a function of $\omega$ for $l=2$ with different values of $r_0$.}
	\label{ABS}
\end{figure}

The absorption cross-section of the black hole corresponding to the transmission coefficient can be computed using the relation \cite{Sanchez:1977si}
\bqn\lb{abs}
\sigma= \sum_{l} \frac{(2l + 1) \pi}{\omega^2} |A|^2.
\eqn
We plot the relation between the absorption cross-section \eqref{abs} and frequency $\omega$ with different values of the parameter $r_0$ in Fig. \ref{ABS}. It shows that the peak of the absorption cross-section increases with the parameter $r_0$.

\begin{figure}[!t]
	\centering
	\includegraphics[scale =0.28]{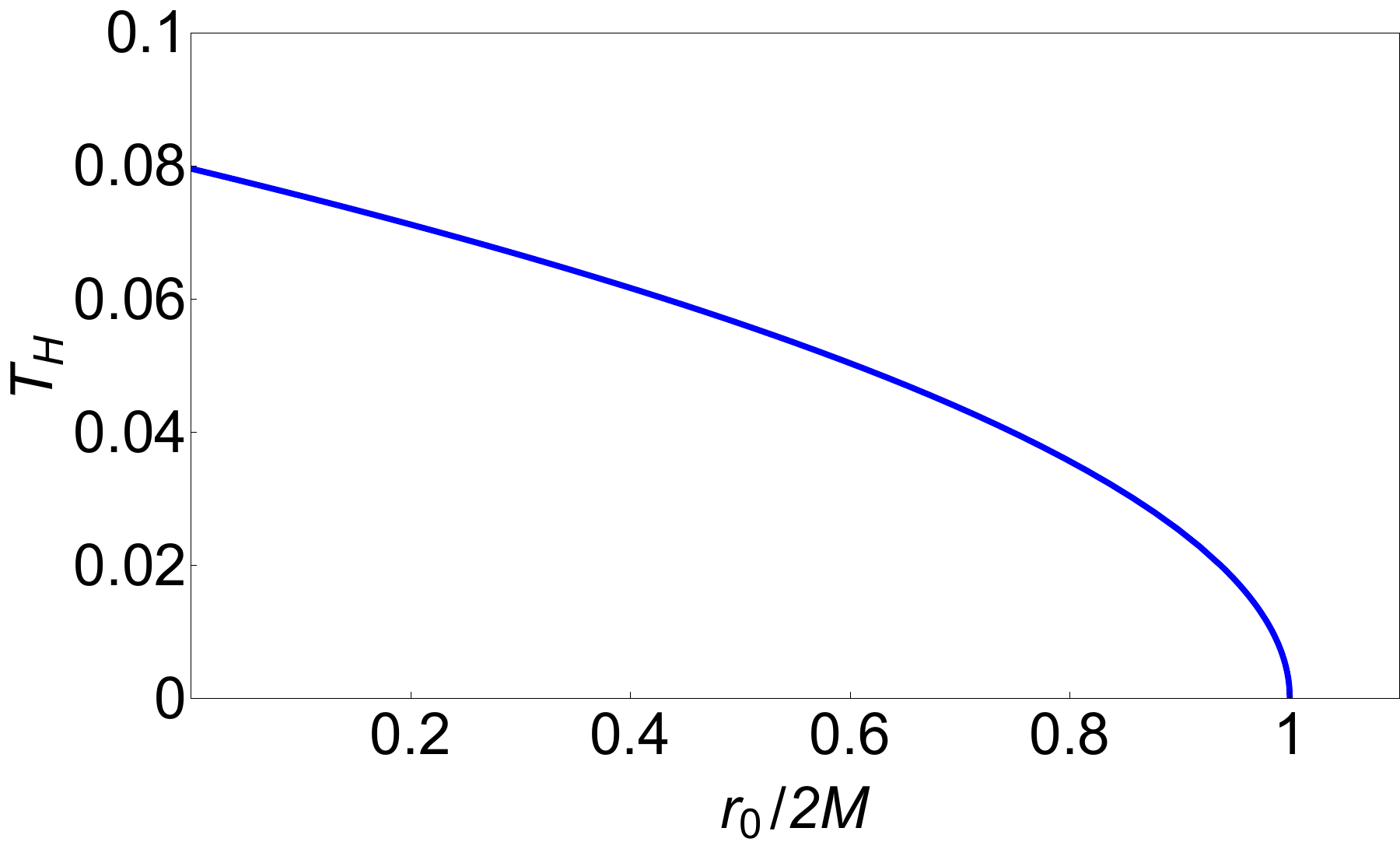}
	\caption{Hawking temperature of the holonomy corrected Schwarzschild black hole as a function of the parameter $r_0$.}
	\label{HawkingT}
\end{figure}

When delving into the properties of black holes, the Hawking temperature undoubtedly emerges as an indispensable concept \cite{Hawking:1974rv}. Situated at the intersection of general relativity and quantum mechanics, the Hawking temperature serves as a pivotal indicator, elucidating the thermodynamic characteristics of microscopic particle motion within black holes. The Hawking temperature of the holonomy-corrected black hole is 
\bqn\lb{HT}
T_{\tx{H}} = \left. \frac{\sqrt{g(r)} \p_{r} f(r)}{4 \pi}  \right|_{r = r_{\tx{H}}}.
\eqn 
We plot the relation between the Hawking temperature \eqref{HT} and the parameter $r_0$ in Fig. \ref{HawkingT}. It shows that the Hawking temperature of the holonomy-corrected black hole decreases with the parameter $r_0$.

\begin{figure}[!t]
	\centering
	\includegraphics[scale =0.28]{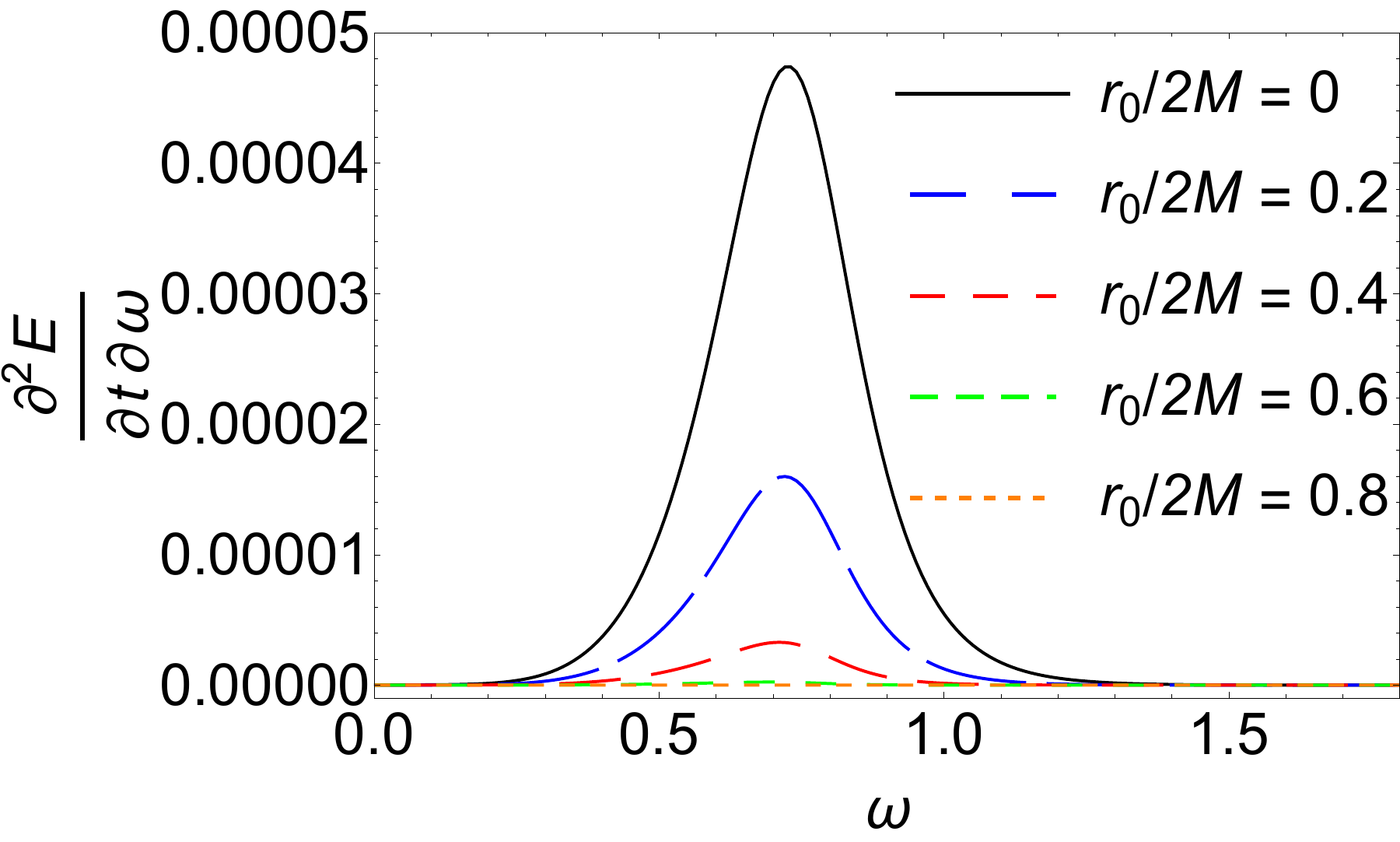}
	\caption{The energy emission rate as a function of $\omega$ for $l=2$ with different values of $r_0$.}
	\label{EER}
\end{figure}

In the pursuit of understanding the intricate dynamics of black holes, one crucial aspect lies in comprehending the energy emission rate of Hawking radiation \cite{Hawking:1975vcx}. This radiation, theorized by Stephen Hawking, fundamentally alters our perception of black holes, revealing their subtle interplay with quantum mechanics and thermodynamics. By calculating the energy emission rate of Hawking radiation for a black hole, we unlock valuable insights into its fundamental properties and behavior. The energy emission rate for Hawking radiation is described as
\bqn\lb{eer}
\frac{dE}{dt} = \sum_l 2(2l+1)|A|^2 \frac{\omega}{e^{\omega/T_{\tx{H}}} -1} \frac{d \omega}{2 \pi}.
\eqn
We plot the relation between the energy emission rate and frequency $\omega$ with different values of the parameter $r_0$ in Fig. \ref{EER}. It shows that the peak of the energy emission rate decreases with the parameter $r_0$.

\section{Conclusions and Discussions}\lb{sec5}

In this work, we studied the QNMs and the greybody factor for the axial gravitational perturbation of the holonomy-corrected Schwarzschild black hole in loop quantum gravity. Assuming that the holonomy-corrected Schwarzschild black hole is described by Einstein’s gravity minimally coupled to an anisotropic perfect fluid, we derived the master equation and effective potential of the axial gravitational perturbation of the holonomy-corrected Schwarzschild black hole. We found that the height of the effective potential decreases with the quantum correction parameter $r_0$, which is consistent with the massless scalar field perturbations of the holonomy corrected Schwarzschild black hole \cite{Fu:2023drp, Moreira:2023cxy}.

With different values of the quantum correction parameter $r_0$, we calculated the QNMs of the axial gravitational perturbation of the holonomy-corrected Schwarzschild black hole by using the WKB approximation method and the asymptotic iteration method. Considering the error in the numerical calculation process, the results obtained from the two different methods show good agreement. We found that the real and imaginary parts of the quasinormal frequencies increase and decrease with the parameter $r_0$, respectively. This agrees with the conclusion for the perturbations of the electromagnetic field and the Dirac field of the holonomy-corrected Schwarzschild black hole \cite{Fu:2023drp, Bolokhov:2023bwm}, but differs from the conclusion for the perturbations of the massless scalar field \cite{Fu:2023drp, Moreira:2023cxy, Bolokhov:2023bwm}. We also found that the influence from the parameter $r_0$ on overtone modes is more perceptible than the influence on fundamental modes. We estimated the numerical errors of the quasinormal frequencies we obtained by different methods. The results show that the errors are very small. Then, we explored the numerical evolution of an initial Gaussian wave packet in the holonomy-corrected Schwarzschild black hole spacetime, and found that the parameter $r_0$ does not significantly affect the late-time power-law behavior of the axial gravitational perturbation of the holonomy-corrected Schwarzschild black hole. This result agrees with the conclusion for the massless scalar field perturbations of the holonomy-corrected Schwarzschild black hole \cite{Moreira:2023cxy}. We derived the parameterized expression of the axial gravitational QNMs of the holonomy-corrected black hole with the parameterized QNMs method of non-rotating black holes proposed by Cardoso \textit{et al.} \cite{Cardoso:2019mqo, McManus:2019ulj}. We found that $r_0 \leq 10^{-2}$ is a necessary condition for the parameterized approximation to be valid. Finally, we calculated the greybody factor for the effective potential of the axial gravitational perturbations of the holonomy-corrected Schwarzschild black hole, and found that when $r_0$ is large, the greybody factor is small for small $\omega$, but it is large for large $\omega$.

During this work, the LIGO–Virgo–KAGRA collaboration started the fourth observing run. It is expected that an increasing number of events of gravitational waves caused by compact binary components will be detected in the future. And one could study the perturbations of black holes and test gravitational theories with more ringdown signals \cite{Berti:2018vdi}.  In realistic astrophysical situations, a tiny perturbation may be added
to the effective potential of the perturbation of a black hole and the quasinormal mode spectrum may be unstable \cite{Cheung:2021bol}. The influence of the quantum correction on the stability of the quasinormal mode spectrum is worth exploring. On the other hand, the framework of parametrized black hole quasinormal ringdown proposed by Cardoso \textit{et~al.} \cite{Cardoso:2019mqo, McManus:2019ulj} may provide us with a theory-agnostic method to study the QNMs of black holes. We will study these issues in future works.

\section*{Acknowledgements}
We thank Xiao-Mei Kuang for important discussions. This work was supported by the National Key Research and Development Program of China (Grant No. 2021YFC2203003), the National Natural Science Foundation of China (Grants No. 12205129, No. 12147166, No. 12247101, and No. 12347111), the China Postdoctoral Science Foundation (Grant No. 2021M701529 and No. 2023M741148), the 111 Project (Grant No. B20063), the Major Science and Technology Projects of Gansu Province, and Lanzhou City’s scientific research funding subsidy to Lanzhou University.

\appendix

\section{The WKB approximation}\lb{AppA}
\renewcommand{\theequation}{A.\arabic{equation}}
\setcounter{equation}{0}

The WKB approximation is a semiclassical method for solving differential equations with spatially varying coefficients. Schutz and Will first applied the WKB approximation to the problem of scattering around a black hole at the first order \cite{Schutz:1985km}. Then the WKB approximation was developed to higher orders \cite{Iyer:1986np, Konoplya:2003ii, Matyjasek:2017psv, Konoplya:2019hlu}. When applied to the study of black hole QNMs, the WKB approximation allows for the calculation of quasinormal frequencies by writing the wave equation governing perturbations in a black hole spacetime as a Schrödinger-like equation \eqref{master-eq}.

The WKB approximation requires that the effective potential of the equation being solved asymptotically approaches constants. The effective potential (\ref{V}) satisfies this condition. The WKB approximation hinges on matching solutions in regions where the effective potential varies slowly (the ``classical turning points") with solutions in the regions where the effective potential is rapidly varying. This matching process leads to a quantization condition, which yields the WKB formula \cite{Konoplya:2003ii}
\bq\lb{WKB}
\f{i V(r_p)}{\sqrt{2 V''(r_p)}} - \sum^N_{i = 2} \Lambda_i = n + \f{1}{2},
\eq 
where $r_p$ is the location of the peak of the effective potential, $n = 0, 1, 2, ...$ is the overtone number, $N$ is the number of the WKB order, and $\Lambda_i$ is the $i$-th correction term. And one can solve Eq. \eqref{WKB} to obtain the quasinormal frequencies.

In this work, we use the 6th-order WKB approximation to calculate the quasinormal frequencies of the axial gravitational perturbation of the holonomy-corrected Schwarzschild black hole. To estimate the numerical error of the quasinormal frequencies obtained from 6th-order WKB approximation, we define the error of the 6th-order WKB approximation as \cite{Konoplya:2019hlu}
\bq\lb{WKB-err}
\Delta_{\tx{WKB}} = \f{|\omega_7 - \omega_5|}{2},
\eq 
where $\omega_7$ and $\omega_5$ are the quasinormal frequencies calculated from 7th and 5th order WKB approximations, respectively.

\section{The asymptotic iteration method}\lb{AppB}
\renewcommand{\theequation}{B.\arabic{equation}}
\setcounter{equation}{0} 

The asymptotic iteration method is a semi-analytic technique for solving eigenvalue problems, $i.e.$ second-order homogeneous linear differential equations \cite{AIM, AIM2}. It is an efficient and accurate technique for calculating QNMs of black hole perturbations and was developed by H.~T.~Cho $et~ al.$ \cite{Cho:2009cj, Cho:2011sf}. To use the asymptotic iteration method to calculate the QNMs of the axial gravitational perturbation of the holonomy-corrected Schwarzschild black hole, we first rewrite the master equation \eqref{master-eq} in the $r$-coordinate
\bqn\lb{master-eq2}
f^2(r) g(r) \Psi''(r) &+& \frac{1}{2} [f^2(r) g(r)]' \Psi'
(r) \nb\\
&+& [\omega^2 - V(r)] \Psi (r) = 0,
\eqn
where prime denotes the derivative with respect to $r$. We rewrite the solution of Eq. \eqref{master-eq2} as
\bqn\lb{AIM-solution}
\Psi(r) &=& \left( 1 - \f{r_H}{r} \right)^{- i \omega / \sqrt{1 - r_0}} \left( \f{r_H}{r} \right)^{- i (2 + r_0) \omega / 2} \nb\\
&& \times e^{\f{i \omega r}{r_H} \sqrt{1 - \f{r_0 r_H}{ r} }} \psi(r),
\eqn
where $\psi(r)$ is a finite and convergent function. Then, we define a new variable $u = 1 - r_H/r$, so $0 \leq u < 1$, and  $u \simeq 1$ at the spatial infinity and $u = 0$ at the event horizon. Using the solution \eqref{AIM-solution} and the new variable $u$, one can rewrite the wave equation \eqref{master-eq2} as the standard form in the asymptotic iteration method
\bqn\lb{AIM1}
\f{d^2 \psi(u)}{ d u^2} = \lambda_0 (u) \f{d \psi(u)}{d u} + s_0 (u) \psi (u),
\eqn
where $\lambda_0 (u)$ and $s_0 (u)$ are polynomial coefficients.
After differentiating Eq. \eqref{AIM1} $n$ times with respect to $u$, one can obtain \cite{ Cho:2011sf}
\bqn
\f{d ^ {n+2} \psi(u)}{d u^{n+2}} = \lambda_n (u) \f{d \psi(u)}{d u} + s_n (u) \psi(u),
\eqn
where
\bqn\lb{AIM2-1}
\lambda_n (u) = \f{d \lambda_{n-1}(u)}{d u} + s_{n-1} (u) + \lambda_0(u) \lambda_{n-1} (u) 
\eqn
and
\bqn\lb{AIM2-2}
s_n (u) = \f{d s_{n-1} (u)}{d u} + s_0 (u) \lambda_{n-1} (u).
\eqn 
Then we could expand $\lambda_n$ and $s_n$ in a Taylor series around the point $\xi$ at which the asymptotic
iteration method is performed:
\bqn\lb{AIM3-1}
\lambda_n (u) = \sum_{i=0}^{\infty} c^i_n (u - \xi)^i
\eqn
and
\bqn\lb{AIM3-2}
s_n (u) = \sum_{i=0}^{\infty} d^i_n (u - \xi)^i,
\eqn 
where $c^i_n$ and $d^i_n$ are the $i$-\text{th} Taylor coefficient’s of $\lambda_n(\xi)$ and $s_n(\xi)$, respectively. Substituting Eqs. \eqref{AIM3-1} and \eqref{AIM3-2} into \eqref{AIM2-1} and \eqref{AIM2-2}, one can get a set of recursion relations for the coefficients
\bqn\lb{AIM4-1}
c^i_n = (i+1) c^{i+1}_{n-1} + d^i_{n-1} + \sum_{k=0}^i c^k_0 c^{i-k}_{n-1}
\eqn
and
\bqn\lb{AIM4-2}
d^i_n = (i+1) c^{i+1}_{n-1} + \sum_{k=0}^i d^k_0 c^{i-k}_{n-1} .
\eqn 
For sufficiently large $n$, the asymptotic aspects of Eqs. \eqref{AIM2-1} and \eqref{AIM2-2} satisfy the following 
quantization condition
\bqn\lb{AIM5}
\f{s_n (u)}{\lambda_n (u)} = \f{s_{n-1} (u)}{\lambda_{n-1} (u)}.
\eqn 
Substituting Eqs. \eqref{AIM3-1}, \eqref{AIM3-2}, \eqref{AIM4-1}, and \eqref{AIM4-2} into  Eq. \eqref{AIM5}, the “quantization condition” can be expressed as 
\bqn\lb{AIM6}
d^0_n c^0_{n-1} - d^0_{n-1} c^0_n = 0.
\eqn
Then, we can obtain the quasinormal frequencies $\omega$ by solving Eq. \eqref{AIM6}. 

In this work, we expand $\lambda_n$ and $s_n$ in a Taylor series up to the 30th order in Eqs. \eqref{AIM3-1} and \eqref{AIM3-2} to calculate the quasinormal frequencies. To estimate the numerical error of the results, we define the error of the asymptotic iteration method with 30th-order expansion as
\bq\lb{AIM-err}
\Delta_{\text{AIM}} = \f{|\omega_{35} - \omega_{25}|}{2},
\eq
where $\omega_{35}$ and $\omega_{25}$ are the quasinormal frequencies calculated from the asymptotic iteration method with 35th-order expansion and with 25th-order expansion, respectively.


\begin{thebibliography}{399}

\bibitem{LIGOScientific:2016aoc}
B.~P.~Abbott \textit{et al.} [LIGO Scientific and Virgo], Observation of Gravitational Waves from a Binary Black Hole Merger,
\href{\doibase 10.1103/PhysRevLett.116.061102}{Phys. Rev. Lett. \textbf{116}, no.6, 061102 (2016)},
[arXiv:1602.03837 [gr-qc]].

\bibitem{LIGOScientific:2016lio}
B.~P.~Abbott \textit{et al.} [LIGO Scientific and Virgo], Tests of general relativity with GW150914,
\href{\doibase 10.1103/PhysRevLett.116.221101}{Phys. Rev. Lett. \textbf{116}, no.22, 221101 (2016)}
[erratum: Phys. Rev. Lett. \textbf{121}, no.12, 129902 (2018)],
[arXiv:1602.03841 [gr-qc]].

\bibitem{LIGOScientific:2018mvr}
B.~P.~Abbott \textit{et al.} [LIGO Scientific and Virgo], GWTC-1: A Gravitational-Wave Transient Catalog of Compact Binary Mergers Observed by LIGO and Virgo during the First and Second Observing Runs, 
\href{\doibase 10.1103/PhysRevX.9.031040}{Phys. Rev. X \textbf{9}, no.3, 031040 (2019)},
[arXiv:1811.12907 [astro-ph.HE]].

\bibitem{LIGOScientific:2020ibl}
R.~Abbott \textit{et al.} [LIGO Scientific and Virgo], GWTC-2: Compact Binary Coalescences Observed by LIGO and Virgo During the First Half of the Third Observing Run,
\href{\doibase 10.1103/PhysRevX.11.021053}{Phys. Rev. X \textbf{11}, 021053 (2021)},
[arXiv:2010.14527 [gr-qc]].

\bibitem{LIGOScientific:2021usb}
R.~Abbott \textit{et al.} [LIGO Scientific and VIRGO], GWTC-2.1: Deep Extended Catalog of Compact Binary Coalescences Observed by LIGO and Virgo During the First Half of the Third Observing Run,
\href{\doibase 10.1103/PhysRevD.109.022001}{Phys. Rev. D \textbf{109}, no.2, 022001 (2024)}, [arXiv:2108.01045 [gr-qc]].

\bibitem{LIGOScientific:2021djp}
R.~Abbott \textit{et al.} [LIGO Scientific, VIRGO and KAGRA], GWTC-3: Compact Binary Coalescences Observed by LIGO and Virgo During the Second Part of the Third Observing Run,
\href{\doibase 10.1103/PhysRevX.13.041039}{Phys. Rev. X \textbf{13}, no.4, 041039 (2023)}, [arXiv:2111.03606 [gr-qc]].

\bibitem{LIGOScientific:2019fpa}
B.~P.~Abbott \textit{et al.} [LIGO Scientific and Virgo], Tests of General Relativity with the Binary Black Hole Signals from the LIGO-Virgo Catalog GWTC-1,
\href{\doibase 10.1103/PhysRevD.100.104036}{Phys. Rev. D \textbf{100}, no.10, 104036 (2019)},
[arXiv:1903.04467 [gr-qc]].

\bibitem{LIGOScientific:2020tif}
R.~Abbott \textit{et al.} [LIGO Scientific and Virgo], Tests of general relativity with binary black holes from the second LIGO-Virgo gravitational-wave transient catalog,
\href{\doibase 10.1103/PhysRevD.103.122002}{Phys. Rev. D \textbf{103}, no.12, 122002 (2021)},
[arXiv:2010.14529 [gr-qc]].

\bibitem{LIGOScientific:2021sio}
R.~Abbott \textit{et al.} [LIGO Scientific, VIRGO and KAGRA],
Tests of General Relativity with GWTC-3,
\href{https://arxiv.org/abs/2112.06861}{[arXiv:2112.06861 [gr-qc]]}.

\bibitem{Chandrasekhar:1985kt}
S.~Chandrasekhar, The mathematical theory of black holes, Oxford University Press, New York, 1983.

\bibitem{M. Maggiore}
M. Maggiore, Gravitational Waves. Vol. 2: Astrophysics and Cosmology, Oxford University Press, 2018, ISBN 978-0-19857089-9.

\bibitem{Kokkotas:1999bd}
K.~D.~Kokkotas and B.~G.~Schmidt, Quasinormal modes of stars and black holes,
\href{\doibase 10.12942/lrr-1999-2}{Living Rev. Rel. \textbf{2}, 2 (1999)},
[arXiv:9909058 [gr-qc]].

\bibitem{Nollert:1999ji}
H.~P.~Nollert,
TOPICAL REVIEW: Quasinormal modes: the characteristic `sound' of black holes and neutron stars,
\href{\doibase 10.1088/0264-9381/16/12/201}{Class. Quant. Grav. \textbf{16}, R159-R216 (1999)}.

\bibitem{Berti:2009kk}
E.~Berti, V.~Cardoso, and A.~O.~Starinets, Quasinormal modes of black holes and black branes,
\href{\doibase 10.1088/0264-9381/26/16/163001}{Class. Quant. Grav. \textbf{26}, 163001 (2009)},
[arXiv:0905.2975 [gr-qc]].

\bibitem{Konoplya:2011qq}
R.~A.~Konoplya and A.~Zhidenko,
Quasinormal modes of black holes: From astrophysics to string theory,
\href{\doibase 10.1103/RevModPhys.83.793}{Rev. Mod. Phys. \textbf{83}, 793-836 (2011)},
[arXiv:1102.4014 [gr-qc]].

\bibitem{Regge:1957td}
T.~Regge and J.~A.~Wheeler,
Stability of a Schwarzschild singularity,
\href{\doibase 10.1103/PhysRev.108.1063}{Phys. Rev. \textbf{108}, 1063-1069 (1957)}.

\bibitem{Zerilli:1970se}
F.~J.~Zerilli,
Effective potential for even parity Regge-Wheeler gravitational perturbation equations,
\href{\doibase 10.1103/PhysRevLett.24.737}{Phys. Rev. Lett. \textbf{24}, 737-738 (1970)}.

\bibitem{Moncrief:1974gw}
V.~Moncrief,
Odd-parity stability of a Reissner-Nordstrom black hole,
\href{\doibase 10.1103/PhysRevD.9.2707}{Phys. Rev. D \textbf{9}, 2707-2709 (1974)}.

\bibitem{Moncrief:1974ng}
V.~Moncrief,
Stability of Reissner-Nordstrom black holes,
\href{\doibase 10.1103/PhysRevD.10.1057}{Phys. Rev. D \textbf{10}, 1057-1059 (1974)}.

\bibitem{Teukolsky:1972my}
S.~A.~Teukolsky,
Rotating black holes - separable wave equations for gravitational and electromagnetic perturbations,
\href{\doibase 10.1103/PhysRevLett.29.1114}{Phys. Rev. Lett. \textbf{29}, 1114-1118 (1972)}.

\bibitem{Liu:2022csl}
W.~Liu, X.~Fang, J.~Jing, and A.~Wang,
Gauge invariant perturbations of general spherically symmetric spacetimes,
\href{\doibase 10.1007/s11433-022-1956-4}{Sci. China Phys. Mech. Astron. \textbf{66}, no.1, 210411 (2023)},
[arXiv:2201.01259 [gr-qc]].

\bibitem{Schutz:1985km}
B.~F.~Schutz and C.~M.~Will,
BLACK HOLE NORMAL MODES: A SEMIANALYTIC APPROACH,
\href{\doibase 10.1086/184453}{Astrophys. J. Lett. \textbf{291}, L33-L36 (1985)}.

\bibitem{Iyer:1986np}
S.~Iyer and C.~M.~Will,
Black Hole Normal Modes: A {WKB} Approach. 1. Foundations and Application of a Higher Order {WKB} Analysis of Potential Barrier Scattering,
\href{\doibase doi:10.1103/PhysRevD.35.3621}{Phys. Rev. D \textbf{35} (1987), 3621}.

\bibitem{Konoplya:2003ii}
R.~A.~Konoplya, Quasinormal behavior of the d-dimensional Schwarzschild black hole and higher order WKB approach, \href{\doibase doi:10.1103/PhysRevD.68.024018}{Phys. Rev. D \textbf{68}, 024018 (2003)},
[arXiv:gr-qc/0303052 [gr-qc]].

\bibitem{Matyjasek:2017psv}
J.~Matyjasek and M.~Opala,
Quasinormal modes of black holes. The improved semianalytic approach,
\href{\doibase doi:10.1103/PhysRevD.96.024011}{Phys. Rev. D \textbf{96}, no.2, 024011 (2017)},
[arXiv:1704.00361 [gr-qc]].

\bibitem{Konoplya:2019hlu}
R.~A.~Konoplya, A.~Zhidenko, and A.~F.~Zinhailo, Higher order WKB formula for quasinormal modes and grey-body factors: recipes for quick and accurate calculations, \href{\doibase doi:10.1088/1361-6382/ab2e25}{Class. Quant. Grav. \textbf{36}, 155002 (2019)},
[arXiv:1904.10333 [gr-qc]].

\bibitem{Leaver:1985ax}
E.~W.~Leaver,
An Analytic representation for the quasi normal modes of Kerr black holes,
\href{\doibase 10.1098/rspa.1985.0119}{Proc. Roy. Soc. Lond. A \textbf{402}, 285-298 (1985)}.

\bibitem{Leaver:1990zz}
E.~W.~Leaver,
Quasinormal modes of Reissner-Nordstrom black holes,
\href{\doibase 10.1103/PhysRevD.41.2986}{Phys. Rev. D \textbf{41}, 2986-2997 (1990)}.

\bibitem{AIM}
Ciftci, Hakan, Richard L. Hall, and Nasser Saad. Asymptotic iteration method for eigenvalue problems, \href{\doibase 
	https://doi.org/10.1088/0305-4470/36/47/008}{Journal of Physics A: Mathematical and General 36.47 (2003): 11807}.

\bibitem{AIM2}
Ciftci, Hakan, Richard L. Hall, and Nasser Saad. Perturbation theory in a framework of iteration methods, \href{\doibase 10.1088/0305-4470/36/47/008}{Physics Letters A 340.5-6 (2005): 388-396}.

\bibitem{Cho:2009cj}
H.~T.~Cho, A.~S.~Cornell, J.~Doukas, and W.~Naylor,
Black hole quasinormal modes using the asymptotic iteration method,
\href{\doibase doi:10.1088/0264-9381/27/15/155004}{Class. Quant. Grav. \textbf{27} (2010), 155004},
[arXiv:0912.2740 [gr-qc]].

\bibitem{Cho:2011sf}
H.~T.~Cho, A.~S.~Cornell, J.~Doukas, T.~R.~Huang, and W.~Naylor, A New Approach to Black Hole Quasinormal Modes: A Review of the Asymptotic Iteration Method,
\href{\doibase doi:10.1155/2012/281705}{Adv. Math. Phys. \textbf{2012}, 281705 (2012)},
[arXiv:1111.5024 [gr-qc]].

\bibitem{Motl:2003cd}
L.~Motl and A.~Neitzke,
Asymptotic black hole quasinormal frequencies,
\href{\doibase 10.4310/ATMP.2003.v7.n2.a4}{Adv. Theor. Math. Phys. \textbf{7}, no.2, 307-330 (2003)},
[arXiv:0301173 [hep-th]].

\bibitem{Horowitz:1999jd}
G.~T.~Horowitz and V.~E.~Hubeny,
Quasinormal modes of AdS black holes and the approach to thermal equilibrium,
\href{\doibase 10.1103/PhysRevD.62.024027}{Phys. Rev. D \textbf{62}, 024027 (2000)},
[arXiv:9909056 [hep-th]].

\bibitem{Berti:2009wx}
E.~Berti, V.~Cardoso, and P.~Pani,
Breit-Wigner resonances and the quasinormal modes of anti-de Sitter black holes,
\href{\doibase 10.1103/PhysRevD.79.101501}{Phys. Rev. D \textbf{79}, 101501 (2009)},
[arXiv:0903.5311 [gr-qc]].

\bibitem{Lin:2016sch}
K.~Lin and W.~L.~Qian,
A Matrix Method for Quasinormal Modes: Schwarzschild Black Holes in Asymptotically Flat and (Anti-) de Sitter Spacetimes,
\href{\doibase 10.1088/1361-6382/aa6643}{Class. Quant. Grav. \textbf{34}, no.9, 095004 (2017)},
[arXiv:1610.08135 [gr-qc]].

\bibitem{Shen:2022xdp}
S.~F.~Shen, W.~L.~Qian, K.~Lin, C.~G.~Shao and Y.~Pan,
Matrix method for perturbed black hole metric with discontinuity,
\href{\doibase 10.1088/1361-6382/ac95f1}{Class. Quant. Grav. \textbf{39}, no.22, 225004 (2022)},
[arXiv:2203.14320 [gr-qc]].

\bibitem{Leung:1997was}
P.~T.~Leung, Y.~T.~Liu, W.~M.~Suen, C.~Y.~Tam, and K.~Young, Quasinormal modes of dirty black holes,
\href{\doibase 10.1103/PhysRevLett.78.2894}{Phys. Rev. Lett. \textbf{78} (1997), 2894-2897}, [arXiv:gr-qc/9903031 [gr-qc]].

\bibitem{Leung:1999iq}
P.~T.~Leung, Y.~T.~Liu, W.~M.~Suen, C.~Y.~Tam, and K.~Young,
Perturbative approach to the quasinormal modes of dirty black holes,
\href{\doibase 10.1103/PhysRevD.59.044034}{Phys. Rev. D \textbf{59} (1999), 044034},[arXiv:gr-qc/9903032 [gr-qc]].

\bibitem{Cardoso:2019mqo}
V.~Cardoso, M.~Kimura, A.~Maselli, E.~Berti, C.~F.~B.~Macedo, and R.~McManus, Parametrized black hole quasinormal ringdown: Decoupled equations for nonrotating black holes, \href{\doibase 10.1103/PhysRevD.99.104077}{Phys. Rev. D \textbf{99}, no.10, 104077 (2019)}, [arXiv:1901.01265 [gr-qc]].

\bibitem{McManus:2019ulj}
R.~McManus, E.~Berti, C.~F.~B.~Macedo, M.~Kimura,~A.~Maselli,~and~V.~Cardoso, Parametrized black hole quasinormal ringdown. II. Coupled equations and quadratic corrections for nonrotating black holes, \href{\doibase doi:10.1103/PhysRevD.100.044061}{Phys. Rev. D \textbf{100} no.4, 044061 (2019)}, [arXiv:1906.05155 [gr-qc]].

\bibitem{Kimura:2020mrh}
M.~Kimura,  Note on the parametrized black hole quasinormal ringdown formalism, \href{\doibase 10.1103/PhysRevD.101.064031}{
Phys. Rev. D \textbf{101} (2020) no.6, 064031},
[arXiv:2001.09613 [gr-qc]].

\bibitem{Hatsuda:2020egs}
Y.~Hatsuda and M.~Kimura, Semi-analytic expressions for quasinormal modes of slowly rotating Kerr black holes, \href{\doibase 10.1103/PhysRevD.102.044032}{Phys. Rev. D \textbf{102} (2020) no.4, 044032},
[arXiv:2006.15496 [gr-qc]].

\bibitem{Volkel:2022aca}
S.~H.~V\"olkel, N.~Franchini, and E.~Barausse, Theory-agnostic reconstruction of potential and couplings from quasinormal modes, \href{\doibase 10.1103/PhysRevD.105.084046}{Phys. Rev. D \textbf{105} (2022) no.8, 084046},
[arXiv:2202.08655 [gr-qc]].

\bibitem{Churilova:2019jqx}
M.~S.~Churilova,
Analytical quasinormal modes of spherically symmetric black holes in the eikonal regime, \href{\doibase 10.1140/epjc/s10052-019-7146-0}{
Eur. Phys. J. C \textbf{79} (2019) no.7, 629},
[arXiv:1905.04536 [gr-qc]].

\bibitem{Hatsuda:2023geo}
Y.~Hatsuda and M.~Kimura,
Perturbative quasinormal mode frequencies,
\href{\doibase 10.1103/PhysRevD.109.044026}{Phys. Rev. D \textbf{109} (2024) no.4, 044026},[arXiv:2307.16626 [gr-qc]].

\bibitem{Franchini:2022axs}
N.~Franchini and S.~H.~V\"olkel, Parametrized quasinormal mode framework for non-Schwarzschild metrics, \href{\doibase 10.1103/PhysRevD.107.124063}{
Phys. Rev. D \textbf{107} (2023) no.12, 124063},[arXiv:2210.14020 [gr-qc]].

\bibitem{Hirano:2024fgp}
S.~Hirano, M.~Kimura, M.~Yamaguchi, and J.~Zhang, Parametrized Black Hole Quasinormal Ringdown Formalism for Higher Overtones,
\href{https://arxiv.org/abs/2404.09672}{[arXiv:2404.09672 [gr-qc]]}.

\bibitem{Bamber:2021knr}
J.~Bamber, O.~J.~Tattersall, K.~Clough, and P.~G.~Ferreira, Quasinormal modes of growing dirty black holes,
\href{\doibase 10.1103/PhysRevD.103.124013}{Phys. Rev. D \textbf{103} (2021) no.12, 124013} [arXiv:2103.00026 [gr-qc]].

\bibitem{Volkel:2022khh}
S.~H.~V\"olkel, N.~Franchini, E.~Barausse, and E.~Berti, Constraining modifications of black hole perturbation potentials near the light ring with quasinormal modes,
\href{\doibase 10.1103/PhysRevD.106.124036}{Phys. Rev. D \textbf{106} (2022) no.12, 124036}, [arXiv:2209.10564 [gr-qc]].

\bibitem{Ashtekar:2004eh}
A.~Ashtekar and J.~Lewandowski,
Background independent quantum gravity: A Status report,
\href{\doibase 10.1088/0264-9381/21/15/R01}{Class. Quant. Grav. \textbf{21}, R53 (2004)},
[arXiv:gr-qc/0404018 [gr-qc]].

\bibitem{Perez:2017cmj}
A.~Perez,
Black Holes in Loop Quantum Gravity,
\href{\doibase 10.1088/1361-6633/aa7e14}{Rept. Prog. Phys. \textbf{80}, no.12, 126901 (2017)},
[arXiv:1703.09149 [gr-qc]].

\bibitem{Bojowald:2020dkb}
M.~Bojowald,
Black-Hole Models in Loop Quantum Gravity,
\href{\doibase 10.3390/universe6080125}{Universe \textbf{6}, no.8, 125 (2020)},
[arXiv:2009.13565 [gr-qc]].

\bibitem{Ashtekar:2023cod}
A.~Ashtekar, J.~Olmedo, and P.~Singh, Regular black holes from Loop Quantum Gravity,
\href{https://arxiv.org/abs/2301.01309}{[arXiv:2301.01309 [gr-qc]]}.

\bibitem{Zhang:2023yps}
X.~Zhang, Loop Quantum Black Hole, \href{\doibase 10.3390/universe9070313}{Universe \textbf{9} (2023) no.7, 313},
[arXiv:2308.10184 [gr-qc]].

\bibitem{Papanikolaou:2023crz}
T.~Papanikolaou,
Primordial black holes in loop quantum gravity: The effect on the threshold, \href{\doibase 10.1088/1361-6382/acd97d}{Class. Quant. Grav. \textbf{40}, no.13, 134001 (2023)}, [arXiv:2301.11439 [gr-qc]].

\bibitem{Santos:2015gja}
V.~Santos, R.~V.~Maluf, and C.~A.~S.~Almeida,
Quasinormal frequencies of self-dual black holes,
\href{\doibase 10.1103/PhysRevD.93.084047}{Phys. Rev. D \textbf{93}, no.8, 084047 (2016)},
[arXiv:1509.04306 [gr-qc]].

\bibitem{Liu:2020ola}
C.~Liu, T.~Zhu, Q.~Wu, K.~Jusufi, M.~Jamil, M.~Azreg-A\"\i{}nou, and A.~Wang,
Shadow and quasinormal modes of a rotating loop quantum black hole,
\href{\doibase 10.1103/PhysRevD.101.084001}{Phys. Rev. D \textbf{101}, no.8, 084001 (2020)},
[erratum: Phys. Rev. D \textbf{103}, no.8, 089902 (2021)]
[arXiv:2003.00477 [gr-qc]].

\bibitem{Daghigh:2020mog}
R.~G.~Daghigh, M.~D.~Green, J.~C.~Morey, and G.~Kunstatter,
Scalar Perturbations of a Single-Horizon Regular Black Hole,
\href{\doibase 10.1103/PhysRevD.102.104040}{Phys. Rev. D \textbf{102}, no.10, 104040 (2020)},
[arXiv:2009.02367 [gr-qc]].

\bibitem{Daghigh:2020fmw}
R.~G.~Daghigh, M.~D.~Green, and G.~Kunstatter,
Scalar Perturbations and Stability of a Loop Quantum Corrected Kruskal Black Hole,
\href{\doibase 10.1103/PhysRevD.103.084031}{Phys. Rev. D \textbf{103}, no.8, 084031 (2021)},
[arXiv:2012.13359 [gr-qc]].

\bibitem{Santos:2021wsw}
J.~S.~Santos, M.~B.~Cruz, and F.~A.~Brito,
Quasinormal modes of a massive scalar field nonminimally coupled to gravity in the spacetime of self-dual black hole,
\href{\doibase 10.1140/epjc/s10052-021-09884-1}{Eur. Phys. J. C \textbf{81}, no.12, 1082 (2021)},
[arXiv:2103.11212 [hep-th]].

\bibitem{Fu:2023drp}
G.~Fu, D.~Zhang, P.~Liu, X.~M.~Kuang, and J.~P.~Wu,
Peculiar properties in quasi-normal spectra from loop quantum gravity effect,
\href{\doibase 10.1103/PhysRevD.109.026010}{Phys. Rev. D \textbf{109}, no.2, 026010 (2024)}, [arXiv:2301.08421 [gr-qc]].

\bibitem{Moreira:2023cxy}
Z.~S.~Moreira, H.~C.~D.~Lima, Junior, L.~C.~B.~Crispino, and C.~A.~R.~Herdeiro,
Quasinormal modes of a holonomy corrected Schwarzschild black hole,
\href{\doibase doi:10.1103/PhysRevD.107.104016}{Phys. Rev. D \textbf{107}, no.10, 104016 (2023)},
[arXiv:2302.14722 [gr-qc]].

\bibitem{Bolokhov:2023bwm}
S.~V.~Bolokhov, Long-lived quasinormal modes and overtones' behavior of the holonomy corrected black holes,
\href{https://arxiv.org/abs/2311.05503}{[arXiv:2311.05503 [gr-qc]]}.

\bibitem{Bouhmadi-Lopez:2020oia}
M.~Bouhmadi-L\'opez, S.~Brahma, C.~Y.~Chen, P.~Chen, and D.~h.~Yeom, A consistent model of non-singular Schwarzschild black hole in loop quantum gravity and its quasinormal modes,
\href{\doibase 10.1088/1475-7516/2020/07/066}{JCAP \textbf{07}, 066 (2020)},
[arXiv:2004.13061 [gr-qc]].

\bibitem{Yang:2023gas}
S.~Yang, W.~D.~Guo, Q.~Tan, and Y.~X.~Liu,
Axial gravitational quasinormal modes of a self-dual black hole in loop quantum gravity,
\href{\doibase 10.1103/PhysRevD.108.024055}{Phys. Rev. D \textbf{108}, no.2, 2 (2023)}, [arXiv:2304.06895 [gr-qc]].

\bibitem{Alonso-Bardaji:2021yls}
A.~Alonso-Bardaji, D.~Brizuela, and R.~Vera,
An effective model for the quantum Schwarzschild black hole,
\href{\doibase 10.1016/j.physletb.2022.137075}{Phys. Lett. B \textbf{829}, 137075 (2022)},
[arXiv:2112.12110 [gr-qc]].

\bibitem{Alonso-Bardaji:2022ear}
A.~Alonso-Bardaji, D.~Brizuela, and R.~Vera,
Nonsingular spherically symmetric black-hole model with holonomy corrections,
\href{\doibase 10.1103/PhysRevD.106.024035}{Phys. Rev. D \textbf{106}, no.2, 2 (2022)},
[arXiv:2205.02098 [gr-qc]].

\bibitem{Junior:2023xgl}
E.~L.~B.~Junior, F.~S.~N.~Lobo, M.~E.~Rodrigues, and H.~A.~Vieira,
Gravitational lens effect of a holonomy corrected Schwarzschild black hole,
\href{\doibase 10.1103/PhysRevD.109.024004}{Phys. Rev. D \textbf{109}, no.2, 2 (2024)}, [arXiv:2309.02658 [gr-qc]].

\bibitem{Soares:2023uup}
A.~R.~Soares, C.~F.~S.~Pereira, R.~L.~L.~Vit\'oria, and E.~M.~Rocha,
Holonomy corrected Schwarzschild black hole lensing,
\href{\doibase 10.1103/PhysRevD.108.124024}{Phys. Rev. D \textbf{108}, no.12, 124024 (2023)},
[arXiv:2309.05106 [gr-qc]].

\bibitem{Chen:2023bao}
R.~T.~Chen, S.~Li, L.~G.~Zhu, and J.~P.~Wu,
Constraints from solar system tests on a covariant loop quantum black hole,
\href{\doibase 10.1103/PhysRevD.109.024010}{Phys. Rev. D \textbf{109}, no.2, 2 (2024)}, [arXiv:2311.12270 [gr-qc]].

\bibitem{Gingrich:2024tuf}
D.~M.~Gingrich, Quasinormal modes of a nonsingular spherically symmetric black hole effective model with holonomy corrections,
\href{https://arxiv.org/abs/2404.04447}{[arXiv:2404.04447 [gr-qc]]}.

\bibitem{Balart:2024rts}
L.~Balart, G.~Panotopoulos, and \'A.~Rinc\'on, Thermodynamics of the quantum Schwarzschild black hole, \href{https://arxiv.org/abs/2404.18804}{[arXiv:2404.18804 [gr-qc]]}.

\bibitem{Chen:2019iuo}
C.~Y.~Chen and P.~Chen,
Gravitational perturbations of nonsingular black holes in conformal gravity,
\href{\doibase 10.1103/PhysRevD.99.104003}{Phys. Rev. D \textbf{99}, no.10, 104003 (2019)},
[arXiv:1902.01678 [gr-qc]].

\bibitem{Gundlach:1993tp}
C.~Gundlach, R.~H.~Price, and J.~Pullin, Late time behavior of stellar collapse and explosions: 1. Linearized perturbations,
\href{\doibase 10.1103/PhysRevD.49.883}{Phys. Rev. D \textbf{49}, 883-889 (1994)},
[arXiv:gr-qc/9307009 [gr-qc]].

\bibitem{Ching:1995tj}
E.~S.~C.~Ching, P.~T.~Leung, W.~M.~Suen, and K.~Young,
Wave propagation in gravitational systems: Late time behavior,
\href{\doibase doi:10.1103/PhysRevD.52.2118}{Phys. Rev. D \textbf{52}, 2118-2132 (1995)},
[arXiv:gr-qc/9507035 [gr-qc]].

\bibitem{basis}
\href{https://centra.tecnico.ulisboa.pt/network/grit/files/ringdown/}{https://centra.tecnico.ulisboa.pt/network/grit/files/ringdown/}.

\bibitem{Konoplya:2019ppy}
R.~A.~Konoplya and A.~F.~Zinhailo,
Hawking radiation of non-Schwarzschild black holes in higher derivative gravity: a crucial role of grey-body factors,
\href{\doibase 10.1103/PhysRevD.99.104060}{Phys. Rev. D \textbf{99}, no.10, 104060 (2019)},
[arXiv:1904.05341 [gr-qc]].

\bibitem{Konoplya:2023moy}
R.~A.~Konoplya and A.~Zhidenko,
Analytic expressions for quasinormal modes and grey-body factors in the eikonal limit and beyond,
\href{\doibase 10.1088/1361-6382/ad0a52}{Class. Quant. Grav. \textbf{40}, no.24, 245005 (2023)},
[arXiv:2309.02560 [gr-qc]].

\bibitem{Sanchez:1977si}
N.~G.~Sanchez, Absorption and Emission Spectra of a Schwarzschild Black Hole,
\href{\doibase 10.1103/PhysRevD.18.1030}{Phys. Rev. D \textbf{18} (1978), 1030}.

\bibitem{Hawking:1974rv}
S.~W.~Hawking, Black hole explosions,
\href{\doibase 10.1038/248030a0}{Nature \textbf{248} (1974), 30-31.}

\bibitem{Hawking:1975vcx}
S.~W.~Hawking, Particle Creation by Black Holes,
\href{\doibase 10.1007/BF02345020}{Commun. Math. Phys. \textbf{43} (1975), 199-220},
[erratum: Commun. Math. Phys. \textbf{46} (1976), 206].

\bibitem{Berti:2018vdi}
E.~Berti, K.~Yagi, H.~Yang, and N.~Yunes,
Extreme Gravity Tests with Gravitational Waves from Compact Binary Coalescences: (II) Ringdown,
\href{\doibase 10.1007/s10714-018-2372-6}{Gen. Rel. Grav. \textbf{50}, no.5, 49 (2018)},
[arXiv:1801.03587 [gr-qc]].

\bibitem{Cheung:2021bol}
M.~H.~Y.~Cheung, K.~Destounis, R.~P.~Macedo, E.~Berti, and V.~Cardoso,
Destabilizing the Fundamental Mode of Black Holes: The Elephant and the Flea,
\href{\doibase 10.1103/PhysRevLett.128.111103}{Phys. Rev. Lett. \textbf{128} (2022) no.11, 111103},
[arXiv:2111.05415 [gr-qc]].




\end{thebibliography}
\end{document}